\documentclass[preprintnumbers,article,amsmath,amssymb,floatfix,10pt,prd,onecolumn,
superscriptaddress,nofootinbib]{revtex4-2}
\usepackage{bm}
\usepackage{amsfonts}
\usepackage{latexsym}
\usepackage[latin1]{inputenc}
\usepackage{graphicx}
\usepackage{amsmath}
\usepackage{palatino}
\usepackage{mathpazo}
\usepackage{textcomp}
\usepackage{float}
\usepackage{booktabs}
\usepackage{dcolumn}
\usepackage{ragged2e}
\usepackage{hyperref}
\hypersetup{colorlinks,citecolor=blue}
\hypersetup{colorlinks=true,linkcolor=red,filecolor=magenta,    urlcolor=blue}
\usepackage{amsmath}
\usepackage{xcolor}
\usepackage{orcidlink}
\usepackage{epsfig}
\usepackage{subfigure}
\usepackage{commath}

\def\jnl@style{\it}
\def\aaref@jnl#1{{\jnl@style#1}}

\def\aaref@jnl#1{{\jnl@style#1}}

\def\aj{\aaref@jnl{AJ}}                   
\def\apj{\aaref@jnl{ApJ}}                 
\def\apjl{\aaref@jnl{ApJ}}                
\def\apjs{\aaref@jnl{ApJS}}               
\def\apss{\aaref@jnl{Ap\&SS}}             
\def\aap{\aaref@jnl{A\&A}}                
\def\aapr{\aaref@jnl{A\&A~Rev.}}          
\def\aaps{\aaref@jnl{A\&AS}}              
\def\mnras{\aaref@jnl{Mon.~Not.~Roy.~Astron.~Soc.}}             
\def\prd{\aaref@jnl{Phys.~Rev.~D}}        
\def\prc{\aaref@jnl{Phys.~Rev.~C}}  
\def\prl{\aaref@jnl{Phys.~Rev.~Lett.}}    
\def\qjras{\aaref@jnl{QJRAS}}             
\def\skytel{\aaref@jnl{S\&T}}             
\def\ssr{\aaref@jnl{Space~Sci.~Rev.}}     
\def\zap{\aaref@jnl{ZAp}}                 
\def\nat{\aaref@jnl{Nature}}              
\def\aplett{\aaref@jnl{Astrophys.~Lett.}} 
\def\apspr{\aaref@jnl{Astrophys.~Space~Phys.~Res.}} 
\def\physrep{\aaref@jnl{Phys.~Rep.}}      
\def\physscr{\aaref@jnl{Phys.~Scr}}       
\def\commat{\aaref@jnl{Comm.~Math.~Phys.}}              
\def\science{\aaref@jnl{Science}}               
\def\cqg{\aaref@jnl{Classical Quant.~Grav.}}            
\def\jpcs{\aaref@jnl{JPCS}}                                     
\def\ijmpd{\aaref@jnl{Int.~J.~Mod.~Phys.~D}}                    
\def\grg{\aaref@jnl{Gen.~Relat.~Gravit.}}               
\def\rpp{\aaref@jnl{Rep.~Prog.~Phys.}}          
\def\npa{\aaref@jnl{Nucl.~Phys.~A}}        
\def\lrr{\aaref@jnl{Living Rev.~Rel.}}                   
\def\jcap{\aaref@jnl{J.~Cosmology Astropart.~Phys.}}    
\def\rmp{\aaref@jnl{Rev.~Mod.~Phys.}}   
\def\epjc{\aaref@jnl{Eur.~Phys.~J.~C}} 
\def\plb{\aaref@jnl{~Phy.~Lett.~B}} 
\def\mpla{\aaref@jnl{Mod.~Phy.~Lett.~A}} 
\def\arxiv{\aaref@jnl{arxiv.org}}

\allowdisplaybreaks[1]

\begin{document}
\color{black}
\title{Non-exotic static spherically symmetric thin-shell wormhole solution in $f(Q,T)$ gravity}

\author{Moreshwar Tayde\orcidlink{0000-0002-3110-3411}}
\email{moreshwartayde@gmail.com}
\affiliation{Department of Mathematics, Birla Institute of Technology and
Science-Pilani,\\ Hyderabad Campus, Hyderabad-500078, India.}

\author{Sayantan Ghosh\orcidlink{0000-0002-3875-0849}}
\email{sayantanghosh.000@gmail.com}
\affiliation{Department of Mathematics, Birla Institute of Technology and Science-Pilani,\\ Hyderabad Campus, Hyderabad-500078, India.}

\author{P.K. Sahoo\orcidlink{0000-0003-2130-8832}}
\email{pksahoo@hyderabad.bits-pilani.ac.in}
\affiliation{Department of Mathematics, Birla Institute of Technology and
Science-Pilani,\\ Hyderabad Campus, Hyderabad-500078, India.}

\date{\today}

\begin{abstract}
In this study, we have conducted an analysis of traversable wormhole solutions within the framework of linear $f(Q, T) = \alpha Q + \beta T$ gravity, ensuring that all the energy conditions hold for the entire spacetime. The solutions presented in this study were derived through a comprehensive analytical examination of the parameter space associated with the wormhole model. This involved considering the exponents governing the redshift and shape functions, as well as the radius of the wormhole throat ($r_0$), the redshift function value at the throat ($\phi_0$), and the model parameters ($\alpha$ and $\beta$). Also, we have established bounds on these free parameters that guarantee the satisfaction of the energy conditions throughout spacetime and have also provided two solutions. Further, we have used the Israel junction condition to see the stability of a thin-shell around the wormhole. We have also calculated the NEC criteria and potential for such a thin-shell and how it varies with the chosen shape function.
\end{abstract}
\maketitle
\section{Introduction}
Black holes (BHs) have been extensively studied and evidence for their existence has been unambiguous \cite{Abbott, Abbottt, Akiyama} established. With the recent discovery of gravity waves, black holes and their study have become an observational astrophysics domain. Meanwhile, the existence of another interesting solution to general relativity (GR ), the wormhole (WH), has not yet been established and its existence is actively debated. This has encouraged researchers to take up this topic and investigate its possible existence in the universe \cite{Khatsymovsky}. The traditional methods used in confirming the existence of black holes cannot be used for wormholes as they are very different structures. So far, wormhole studies have remained theoretical, though researchers are trying to find more efficient ways for its detection. Nevertheless, it will be interesting to understand if wormholes exist, especially considering their intriguing properties such as time travel, etc. Black holes and wormholes are two distinct entities with a fundamental difference between them. While black hole formation is generic, omnipresent in nature, and requires no special matter content or energy condition (EC); the formation of a wormhole is not as simple or generic. The matter content of a wormhole needs to be special, non-trivial, and exotic, and it must violate the null energy condition (NEC) for it to be formed. This shows that while black holes can be formed under any energy conditions; a wormhole needs to have specific conditions and matter content for its formation \cite{Visser}. Hence, it can be concluded that the formation of a wormhole is more demanding than that of a black hole.

The work done on wormholes received a major boost after the discovery of black holes and the possibility of probing their interior using gravity waves. The concept of wormholes and the Einstein-Rosen bridge were once thought of as only a formal mathematical result \cite{Einstein}. However, Wheeler's pioneering work \cite{Wheeler} emphasized that wormholes could be used to form bridges between two widely separated regions. This opened an entirely new line of research that could be used to study and explore the possibilities of time travel and shortcuts across galaxies. In 1988, Morris and Thorne \cite{Morris} observed another kind of wormhole solution that can be traversed by preserving the entrance to the wormhole. The throat of this wormhole is held open by a special type of matter that violates energy conditions, in particular the NEC. This type of matter is known as Exotic Matter, which is not part of the Standard Model of Particle Physics and is an extension of the Standard Model. Over the years, many models for traversable wormholes have been proposed, including those based on exotic matter \cite{Visser}. Some theories suggest that traversable wormholes may offer an alternative method for interstellar travel, while others propose they can act as gateways to other universes or even parallel universes. Theoretically speaking, the idea of using a wormhole to form a bridge between two distant points is plausible, however, there are still many unknowns. If one was to successfully use a traversable wormhole, a variety of challenges must first be addressed, such as the stability of the throat.

 In recent years, there has been growing interest among researchers in modified theories of gravity, which are the geometrical extensions of Einstein's General Relativity (GR). These theories are used to explain the early and late-time acceleration of the universe, and much work has already been done on astrophysical objects such as wormholes in modified theories. Wormholes, in particular, have sparked a significant amount of research, with the Rastall theory \cite{Moradpour}, Born-Infield theory \cite{Eiroa, Richarte, Shaikh}, curvature matter coupling \cite{Ditta, Garcia, Garcia1}, quadratic gravity \cite{Duplessis}, braneworld \cite{Lobo2, Camera, Bronnikov2, Kar1, Parsaei1}, and  Einstein-Cartan gravity \cite{Bronnikov1, Bronnikov, Mehdizadeh} all being explored to further our understanding of them.  
 Lobo and Oliveira \cite{Miguel} have extensively studied wormhole geometries in the context of $f(R)$ gravity and have used specific shape functions and various equations of state to find the exact wormhole solutions. Further, Azizi \cite{Azizi} studied wormholes in $f(R, T)$ gravity and showed that this violation of the NEC was due to effective stress energy. For furthermore study of wormholes in modified theories of gravity, we can check Refs. \cite{1, Yousaf/2017, P.K. Sahoo, Iqra, Elizalde, Shamaila, Jamil/2013, Harko/2012, Sharif/2017, Sharif/2018}

 The recently introduced symmetric teleparallel gravity or $f(Q)$ gravity where $Q$ represents non-metricity scalar, is a novel class of modified gravity where torsion and curvature are absent. The affine connection plays an important role in this theory as compared to physical manifold \cite{Jimenez}. Additionally, another prominent feature is that the field equations of $f(Q)$ gravity are second order, while in $f(R)$ gravity, they are fourth-order \cite{Faraoni}, making it relatively easier to solve the equations. Nowadays, the non-metricity types of gravity becomes popular among researchers. The observational study on $f(Q)$ gravity has been done in \cite{R. Solanki 1, R. Solanki 2, Maurya, D'Agostino, Lymperis}. Also, in the field of astrophysical objects, we could find some interesting kinds of literature, for instance, blackholes \cite{F. D'Ambrosio, A. Chanda}, wormhole geometries \cite{Z. Hassan 1, Z. Hassan 2, Z. Hassan 3, Z. Hassan 4}, gavastar \cite{S. Pradhan}, spherically symmetric configurations \cite{R. Lin}, quintessence field \cite{S. Mandal} in $f(Q)$ gravity.

 The extension of $f(Q)$ gravity known as $f(Q, T)$ gravity, which is based on the coupling of non-metricity $Q$ and the trace of energy-momentum tensor $T$, was presented by Yixin, et al. \cite{Y. Xu et al} According to this theory, the energy-momentum tensor $T$ serves as a link between the gravitational impact and manifestations from the quantum field. As it was only recently conceived, this gravity has been the subject of much research based on theoretical \cite{Najera, Bhattacharjee} and observational \cite{G. Gadbail} characteristics. In \cite{T. Harko}, Harko, et al. examined the innovative couplings between non-metricity and matter and also talked about coupling matter in modified $Q$ gravity \cite{Harko}.

In the area of non-minimal matter-curvature coupling, Garcia and Lobo \cite{Garcia and Lobo} created several precise WH models and concluded that non-minimal coupling could help to reduce the violation of NEC of the typical WH throat matter content. When the radial component of pressure is proportional to a real constant value of the torsion scalar, theoretical occurrence of WH geometries may be possible, according to Daouda, et al.'s \cite{Daouda et al} investigation of spherically symmetric WH solutions coupled with anisotropic exotic matter content in the context of $f(T)$ gravity. WH models obeying EC at its throat are feasible with specific selections of redshift, $f(T)$ functions, and the shape function in modified gravity, according to B\"ohmer, et al. \cite{Bohmer}. Using a specific shape function and redshift function, Rosa, et al. \cite{Rosa} examined traversable wormhole solutions with the linear form of $f (R, T)$ gravity fulfilling all the respective energy conditions for the entire spacetime.
Furthermore, for references on more studies, one can refer to the articles \cite{Jamil, T Harko et al, Yousaf/2017, Moraes et al}. 


 The study of thin-shell around wormhole was first done by Poisson \cite{E. Poisson}. 
 It was motivated by the earlier work by Poisson \cite{P. R. Brady} 
 where they considered the stability of a thin-shell around the black hole. The idea was to consider a massive but negligible thickness spherically symmetric shell between two different metrics and then use the junction condition given by Israel \cite{W. Israel}
 to calculate the stress and pressure across the shell. From the conservation of the energy-momentum equation, one can find the potential for the shell. The thin-shell across the wormhole has been studied in other gravity models like $f(R)$ gravity \cite{A. Eid}, 
 rastall gravity \cite{I. P. Lobo},  
 gauss bonnet gravity \cite{N. Rahman}, 
 etc. just to name a few. So far best of our knowledge there has been no work done in $f(Q, T)$
 gravity about the stability of the thin-shell around the wormhole. In this article, we have studied a spherically symmetric massive shell of negligible thickness surrounding the wormhole. We have used the junction condition where inside we have taken the wormhole metric, and outside the shell, we have taken the Schwarzschild metric. Following the junction conditions, we get the stress($\sigma$) and pressure($p$) for such thin-shell. From $\sigma$ and $p$ we can get the effective pressure following the prescription of \cite{E. Poisson}. 
The potential $v(r)$ can give some phenomenological description for the deflection of light and ISCO (Inner most stable circular orbit) for the accretion disk \cite{M. Sharif, S. D. Forghani}. 

In this manner, the article's summary is structured: The fundamental theory of $f(Q, T)$ gravity has been demonstrated in Sec. \ref{sec2}, and the associated field equations for a wormhole solution in $f(Q, T)$ are provided in Sec. \ref{sec3}. By considering the specific form of shape function, non-constant redshift function, and a linear form of $f(Q, T)$, we analyze the parameter space of the model under consideration in order to derive the necessary parameter restrictions for the energy conditions to be met and present a few examples of solutions in Sec. \ref{sec4} and also study for the non-linear case of $f(Q, T)$ in \ref{sec5}. We study the junction conditions of the model under consideration in Sec. \ref{sec6} followed by final remarks in Sec. \ref{sec7}.
 
\section{Basic Field Equations in $f(Q, T)$ Gravity}
\label{sec2}
We consider the action for symmetric teleparallel gravity proposed by Y. Xu et al. \cite{Y. Xu et al},
\begin{equation}\label{1}
\mathcal{S}=\int\frac{1}{16\pi}\,f(Q,T)\sqrt{-g}\,d^4x+\int \mathcal{L}_m\,\sqrt{-g}\,d^4x\, ,
\end{equation}
where arbitrary $f$ is a function of non-metricity scalar $Q$ and trace of the energy momentum tensor $T$, $g$ is the determinant of the metric $g_{\mu\nu}$, and $\mathcal{L}_m$ is the matter Lagrangian density.\\
The non-metricity tensor is explicitly given by \cite{Jimenez}\\
\begin{equation}\label{2}
Q_{\lambda\mu\nu}=\bigtriangledown_{\lambda} g_{\mu\nu},
\end{equation}
Additionally, we can define the superpotential or non-metricity conjugate as
\begin{equation}\label{4}
P^\alpha\;_{\mu\nu}=\frac{1}{4}\left[-Q^\alpha\;_{\mu\nu}+2Q_{(\mu}\;^\alpha\;_{\nu)}+Q^\alpha g_{\mu\nu}-\tilde{Q}^\alpha g_{\mu\nu}-\delta^\alpha_{(\mu}Q_{\nu)}\right],
\end{equation}
where traces of the non-metricity tensor are
\begin{equation}
\label{3}
Q_{\alpha}=Q_{\alpha}\;^{\mu}\;_{\mu},\; \tilde{Q}_\alpha=Q^\mu\;_{\alpha\mu}.
\end{equation}
By taking the following contraction from the previous definition, one can deduce the non-metricity scalar as  \cite{Jimenez}
\begin{eqnarray}
\label{5}
Q &=& -Q_{\alpha\mu\nu}\,P^{\alpha\mu\nu}\\
&=& -g^{\mu\nu}\left(L^\beta_{\,\,\,\alpha\mu}\,L^\alpha_{\,\,\,\nu\beta}-L^\beta_{\,\,\,\alpha\beta}\,L^\alpha_{\,\,\,\mu\nu}\right),
\end{eqnarray}
where the disformation $L^\beta_{\,\,\,\mu\nu}$ is described as
\begin{equation}
L^\beta_{\,\,\,\mu\nu}=\frac{1}{2}Q^\beta_{\,\,\,\mu\nu}-Q_{(\mu\,\,\,\,\,\,\nu)}^{\,\,\,\,\,\,\beta}.
\end{equation}

The gravitational equations of motion can now be obtained by varying the action concerning the metric tensor $g_{\mu\nu}$, and they can be written as
\begin{equation}\label{7}
\frac{-2}{\sqrt{-g}}\bigtriangledown_\alpha\left(\sqrt{-g}\,f_Q\,P^\alpha\;_{\mu\nu}\right)-\frac{1}{2}g_{\mu\nu}f \\
+f_T \left(T_{\mu\nu} +\Theta_{\mu\nu}\right)\\
-f_Q\left(P_{\mu\alpha\beta}\,Q_\nu\;^{\alpha\beta}-2\,Q^
{\alpha\beta}\,\,_{\mu}\,P_{\alpha\beta\nu}\right)=8\pi T_{\mu\nu},
\end{equation}

where $f_Q=\frac{\partial f}{\partial Q}$ and $f_T=\frac{\partial f}{\partial T}$.

By definition, the energy-momentum tensor for the fluid depiction of spacetime can be made as
\begin{equation}\label{6}
T_{\mu\nu}=-\frac{2}{\sqrt{-g}}\frac{\delta\left(\sqrt{-g}\,\mathcal{L}_m\right)}{\delta g^{\mu\nu}},
\end{equation}
and
\begin{equation}\label{6a}
\Theta_{\mu\nu}=g^{\alpha\beta}\frac{\delta T_{\alpha\beta}}{\delta g^{\mu\nu}}.
\end{equation}

\section{Wormhole in $f(Q,T)$ Gravity}
\label{sec3}
 Taking into account the static and spherically symmetric wormhole metric with Schwarzschild coordinates $(t,\,r,\,\theta,\,\Phi)$ is given by \cite{Morris, Visser}
\begin{equation}\label{8}
ds^2=e^{2\phi(r)}dt^2-\left(1-\frac{b(r)}{r}\right)^{-1}dr^2-r^2\,d\theta^2-r^2\,\sin^2\theta\,d\Phi^2,
\end{equation}
where $\phi(r)$ and $b(r)$ denote the redshift function and the shape function, respectively. Furthermore, both conform to the following requirements \cite{Morris, Visser}:
\begin{itemize}
  \item[(1)]  For the condition $r>r_0$, the shape function $b(r)$ must satisfy $b(r) < r$. At the throat of the wormhole, where $r=r_0$, the condition $b(r_0)=r_0$ must be met.
  \item[(2)] The shape function $b(r)$ must satisfy the flaring-out condition at the throat, which requires that $b'(r_0) < 1$.
  \item[(3)] The asymptotic flatness condition requires that the limit $\frac{b(r)}{r} \rightarrow 0$ as $r\rightarrow \infty$.
  \item[(4)] Redshift function $\phi(r)$ should be finite everywhere.
\end{itemize}
In the current research, to analyze the wormhole solution, we assume that an anisotropic energy-momentum tensor that is provided by \cite{Morris, Visser}
\begin{equation}\label{10}
T_{\mu}^{\nu}=\left(\rho+p_t\right)u_{\mu}\,u^{\nu}-p_t\,\delta_{\mu}^{\nu}+\left(p_r-p_t\right)v_{\mu}\,v^{\nu},
\end{equation}
where, $\rho$ indicates the energy density, $p_r$ and $p_t$ denote the radial and tangential pressures, respectively and both are a function of radial coordinate $r$. $u_{\mu}$ and $v_{\mu}$ are the four-velocity vector and unitary space-like vectors, respectively. Additionally, both meet the requirements $u_{\mu}u^{\nu}=-v_{\mu}v^{\nu}=1$. It turns out that the energy-momentum tensor's trace is $T=\rho-p_r-2p_t$.\\
In this article, we discuss matter Lagrangian $\mathcal{L}_m=-P$ \cite{Correa,Sergio Mendoza,T. Harko 1,V. M. C. Ferreira} and hence Eq. \eqref{6a} can be viewed
\begin{equation}
    \Theta_{\mu\nu}=-g_{\mu\nu}\,P-2\,T_{\mu\nu},
\end{equation}
where $P=\frac{p_r+2\,p_t}{3}$ represents the total pressure.\\
For the metric \eqref{8}, the non-metricity scalar $Q$ is given by
\begin{equation}\label{9}
Q=-\frac{b}{r^2}\left[\frac{rb^{'}-b}{r(r-b)}+2\phi^{'}\right].
\end{equation}\\
Now, by adding Eqs. \eqref{8}, \eqref{10}, and \eqref{9} into equation of motion \eqref{7}, following are the field equations for $f(Q,T)$ gravity \cite{Tayde}
\begin{equation}\label{11}
8 \pi  \rho =\frac{(r-b)}{2 r^3} \left[f_Q \left(\frac{(2 r-b) \left(r b'-b\right)}{(r-b)^2}+\frac{b \left(2 r \phi '+2\right)}{r-b}\right)+\frac{2 b r f_{\text{QQ}} Q'}{r-b}+\frac{f r^3}{r-b}-\frac{2r^3 f_T (P+\rho )}{(r-b)}\right],
\end{equation}
\begin{equation}\label{12}
8 \pi  p_r=-\frac{(r-b)}{2 r^3} \left[f_Q \left(\frac{b }{r-b}\left(\frac{r b'-b}{r-b}+2 r \phi '+2\right)-4 r \phi '\right)+\frac{2 b r f_{\text{QQ}} Q'}{r-b}+\frac{f r^3}{r-b}-\frac{2r^3 f_T \left(P-p_r\right)}{(r-b)}\right],
\end{equation}
\begin{equation}\label{13}
8 \pi  p_t=-\frac{(r-b)}{4 r^2} \left[f_Q \left(\frac{\left(r b'-b\right) \left(\frac{2 r}{r-b}+2 r \phi '\right)}{r (r-b)}+\frac{4 (2 b-r) \phi '}{r-b}-4 r \left(\phi '\right)^2-4 r \phi ''\right)-4 r f_{\text{QQ}} Q' \phi '+\frac{2 f r^2}{r-b}-\frac{4r^2 f_T \left(P-p_t\right)}{(r-b)}\right].
\end{equation}\\
By taking into account various models of $f(Q, T)$ gravity, one can investigate wormhole solutions using these field equations.\\

Let's spend a few lines discussing the Raychaudhuri equations-derived classical energy conditions. The physically realistic matter configuration is discussed using these conditions. The four energy conditions that are commonly used in general relativity are:\\
$\bullet$ Null energy condition (NEC) which requires $\rho+p_j\geq0$ for all $j$.\\
$\bullet$ Weak energy condition (WEC) which requires $\rho\geq0$ and $\rho+p_j\geq0$ for all $j$.\\
$\bullet$ Strong energy condition (SEC) which requires $\rho+p_j\geq0$ and $\rho+\sum_jp_j\geq0$ for all $j$.\\
$\bullet$ Dominant energy condition (DEC) which requires $\rho\geq0$ and $\rho \pm p_j\geq0$ for all $j$,\\
where $j=r,\, t$.\\
These energy conditions are important in studying the properties of spacetime and the matter sources that generate it. For example, the violation of the NEC in a wormhole solution would indicate the presence of exotic matter in the throat of the wormhole. Additionally, a positive energy density is required for a physically realistic matter source that can sustain a wormhole solution in general relativity.\\
\section{Wormhole solutions with linear $f(Q,T)$}\label{sec4}
In this part, We will examine wormhole solutions with linear functional forms of $f(Q, T)$ gravity which is given by \cite{Y. Xu et al}
\begin{equation}
\label{14}
f(Q,T)=\alpha\,Q+\beta\,T,
\end{equation}
 where $\alpha$ and $\beta$ are model parameters.\\
 Also, we choose redshift function $\phi(r)$ and shape function $b(r)$ as \cite{Bruno}
 \begin{equation}\label{15}
 \phi(r) =\phi _0 \left(\frac{r_0}{r}\right)^{\lambda },
 \end{equation}
 \begin{equation}\label{16}
 b(r)=r_0 \left(\frac{r_0}{r}\right)^{\eta },
 \end{equation}
 respectively, where $\lambda$ and $\eta$ are constant exponents which are strictly positive to satisfy the asymptotic flatness condition. Also, $\phi_0$ is an arbitrary constant. Here we especially consider $\eta > 1$ for analyzing the wormhole solution easier.\\
 Using a linear form of $f(Q, T)$, a particular form of redshift and shape function,  Eqs. \eqref{11} to \eqref{13} gives
 \begin{multline}\label{17}
 \rho =\frac{\alpha}{12 (4 \pi -\beta ) (\beta +8 \pi ) r^3} \left[r_0 \left(\frac{r_0}{r}\right)^{\eta } \left(4 (\beta -12 \pi ) \eta +\lambda  \phi _0 \left(\frac{r_0}{r}\right)^{\lambda } \left(\beta  (5 \eta +10 \lambda -11)+10 \beta  \lambda  \phi _0 \left(\frac{r_0}{r}\right)^{\lambda }-48 \pi \right)\right)  \right.\\\left.
 -10 \beta  \lambda  r \phi _0 \left(\frac{r_0}{r}\right)^{\lambda } \left(\lambda +\lambda  \phi _0 \left(\frac{r_0}{r}\right)^{\lambda }-1\right)\right],
 \end{multline}
 \begin{multline}\label{18}
 p_r=\frac{\alpha}{12 (4 \pi -\beta ) (\beta +8 \pi ) r^3} \left[r_0 \left(\frac{r_0}{r}\right)^{\eta } \left(4 \beta  (2 \eta +3)-\lambda  \phi _0 \left(\frac{r_0}{r}\right)^{\lambda } \left(\beta  (5 \eta +10 \lambda +13)+10 \beta  \lambda  \phi _0 \left(\frac{r_0}{r}\right)^{\lambda }-144 \pi \right)-48 \pi \right)  \right.\\\left.
+2 \lambda  r \phi _0 \left(\frac{r_0}{r}\right)^{\lambda } \left(5 \beta  \lambda +7 \beta +5 \beta  \lambda  \phi _0 \left(\frac{r_0}{r}\right)^{\lambda }-48 \pi \right)\right],
 \end{multline}
 \begin{multline}\label{19}
 p_t=\frac{\alpha}{12 (4 \pi -\beta ) (\beta +8 \pi ) r^3} \left[r_0 \left(\frac{r_0}{r}\right){}^{\eta } \left(2 \beta  (\eta -3)+24 \pi  (\eta +1)+\lambda  \phi _0 \left(\frac{r_0}{r}\right){}^{\lambda } \left(\beta  (\eta +2 \lambda +17)-24 \pi  (\eta +2 \lambda -1)  \right.\right.\right.\\\left.\left.\left.
 -2 (24 \pi -\beta ) \lambda  \phi _0 \left(\frac{r_0}{r}\right){}^{\lambda }\right)\right)+2 \lambda  r \phi _0 \left(\frac{r_0}{r}\right){}^{\lambda } \left(-\beta  (\lambda +5)+24 \pi  \lambda +(24 \pi -\beta ) \lambda  \phi _0 \left(\frac{r_0}{r}\right){}^{\lambda }\right)\right].
 \end{multline}
 So, taking the density and pressures into account from eqs. \eqref{17}-\eqref{19}, we get
 \begin{equation}\label{20}
 \rho+p_r= \frac{-\alpha}{(\beta +8 \pi ) r^3} \left[r_0 \left(\frac{r_0}{r}\right)^{\eta } \left(\eta -2 \lambda  \phi _0 \left(\frac{r_0}{r}\right)^{\lambda }+1\right)+2 \lambda  r \phi _0 \left(\frac{r_0}{r}\right)^{\lambda }\right],
 \end{equation}
 \begin{equation}\label{21}
\rho+p_t= \frac{\alpha}{2 (\beta +8 \pi ) r^3} \left[r_0 \left(\frac{r_0}{r}\right)^{\eta } \left(-\eta -\lambda  \phi _0 \left(\frac{r_0}{r}\right)^{\lambda } \left(\eta +2 \lambda +2 \lambda  \phi _0 \left(\frac{r_0}{r}\right)^{\lambda }+1\right)+1\right)+2 \lambda ^2 r \phi _0 \left(\frac{r_0}{r}\right)^{\lambda } \left(\phi _0 \left(\frac{r_0}{r}\right)^{\lambda }+1\right)\right],
 \end{equation}
 \begin{multline}\label{22}
\rho-p_r= \frac{\alpha}{6 (4 \pi -\beta ) (\beta +8 \pi ) r^3} \left[r_0 \left(\frac{r_0}{r}\right)^{\eta } \left(-2 \beta  (\eta +3)-24 \pi  (\eta -1)+\lambda  \phi _0 \left(\frac{r_0}{r}\right)^{\lambda } \left(5 \beta  \eta +10 \beta  \lambda +\beta +10 \beta  \lambda  \phi _0 \left(\frac{r_0}{r}\right)^{\lambda } \right.\right.\right.\\\left.\left.\left.
-96 \pi \right)\right)-2 \lambda  r \phi _0 \left(\frac{r_0}{r}\right)^{\lambda } \left(5 \beta  \lambda +\beta +5 \beta  \lambda  \phi _0 \left(\frac{r_0}{r}\right)^{\lambda }-24 \pi \right)\right],
 \end{multline}
 \begin{multline}\label{23}
 \rho-p_t=\frac{\alpha}{6 (4 \pi -\beta ) (\beta +8 \pi ) r^3} \left[r_0 \left(\frac{r_0}{r}\right)^{\eta } \left(\beta  (\eta +3)-12 \pi  (3 \eta +1)+2 \lambda  \phi _0 \left(\frac{r_0}{r}\right)^{\lambda } \left(\beta  (\eta +2 \lambda -7)+6 \pi  (\eta +2 \lambda -3) \right.\right.\right.\\\left.\left.\left.
+2 (\beta +6 \pi ) \lambda  \phi _0 \left(\frac{r_0}{r}\right)^{\lambda }\right)\right)+2 \lambda  r \phi _0 \left(\frac{r_0}{r}\right)^{\lambda } \left(-2 (\beta +6 \pi ) \lambda +5 \beta -2 (\beta +6 \pi ) \lambda  \phi _0 \left(\frac{r_0}{r}\right)^{\lambda }\right)\right],
 \end{multline}
 \begin{multline}\label{24}
 \rho+p_r+2 p_t=\frac{\alpha}{6 (4 \pi -\beta ) (\beta +8 \pi ) r^3} \left[r_0 \left(\frac{r_0}{r}\right)^{\eta } \left(8 \beta  \eta +\lambda  \phi _0 \left(\frac{r_0}{r}\right)^{\lambda } \left(\beta  (\eta +2 \lambda +5)-24 \pi  (\eta +2 \lambda -3) \right.\right.\right.\\\left.\left.\left.
 -2 (24 \pi -\beta ) \lambda  \phi _0 \left(\frac{r_0}{r}\right)^{\lambda }\right)\right)+2 (24 \pi -\beta ) \lambda  r \phi _0 \left(\frac{r_0}{r}\right)^{\lambda } \left(\lambda +\lambda  \phi _0 \left(\frac{r_0}{r}\right)^{\lambda }-1\right)\right].
 \end{multline}

Now, using the redshift function $\phi(r)$ and the shape function $b(r)$ given in Eqs. \eqref{15} and \eqref{16} respectively, one can verify that the wormhole solution satisfies energy conditions everywhere. We will discuss it in the next subsections \ref{subsec1} to \ref{subsec5}.

\subsection{Conditions for NEC}\label{subsec1}
Let's start with the analysis of NEC. Taking into consideration the particular form of the redshift function $\phi(r)$ and the shape function $b(r)$ given in Eqs. \eqref{15} and \eqref{16} respectively, one could use the following boundary condition at the throat $r=r_0$ and which is given by
\begin{equation}\label{25}
\rho(r)+p_r(r) \bigg\vert_{(r=\,r_0)} =-\frac{\alpha  (\eta +1)}{(\beta +8 \pi ) {r_0}^2} \,,
\end{equation}
\begin{equation}\label{26}
\rho(r)+p_t(r) \bigg\vert_{(r=\,r_0)} =-\frac{\alpha  \left((\eta +1) \lambda  \phi _0+\eta -1\right)}{2 (\beta +8 \pi ) {r_0}^2} \,.
\end{equation}
As $\eta>0$, one can verify from Eq. \eqref{25} that when $\alpha=1$ and $\beta=0$ that is in the GR case, $\rho+p_r$ is always negative at the throat $r=r_0$, this gives violation of NEC. However, in the general case i.e. $\alpha\neq1$ or $\beta\neq0$, one can verify that $\rho+p_r>0$ along with Eq. \eqref{25}, impose  a constraint on parameter $\alpha$ and $\beta$ i.e.
\begin{equation}\label{27}
\alpha <0,\,\, \beta >-8 \pi \,\,\,\,\text{or} \,\,\,\, \alpha >0,\,\, \beta <-8 \pi\,.
\end{equation} 
In addition, $\rho+p_t>0$ along with Eq. \eqref{26}, impose  a constraint on parameter $\phi_0$ i.e.
\begin{equation}\label{28}
\phi _0>\frac{1-\eta }{\lambda  (\eta+1) }\equiv \phi_c \,.
\end{equation}
If these constraints from Eqs. \eqref{27} and \eqref{28} are satisfied, then NEC will be satisfied at the throat, which is unattainable in GR. Also from Eq. \eqref{28}, one could observe that $\phi_c\rightarrow+\infty$ as $\lambda\rightarrow0$ with $\eta<1$ i.e. NEC is not satisfied for any value of $\phi_0$ at the throat $r=r_0$ and further $\eta =1$ gives $\phi_c=0$. Furthermore, one can notice that $\phi_c$ is undefined as $\lambda \rightarrow 0$ and $\eta \rightarrow 1$. Thus, for defining $\phi_c$, one must restrict $\eta \nrightarrow 1$ if $\lambda \rightarrow 0$. So depending on the combination of the parameters, either NEC is satisfied for the whole spacetime or NEC is satisfied for the finite range of the radial coordinate $r$ around the throat, say $r<r_c$, but it is violated elsewhere for $r>r_c$. \\

To guarantee the physical relevance of the obtained wormhole solutions, it is not enough to satisfy the NEC at the throat. Thus, to guarantee the physical relevance of the obtained wormhole solutions for the whole spacetime, we will start analysis for the combination $\rho+p_r$ and $\rho+p_t$ separately and impose constraints on the parameters $\lambda$, $\eta$, and $\phi_0$ which will guarantee for $\rho+p_r>0$ and $\rho+p_t>0$. Then we will combine the results into a unified set of constraints. 

\subsubsection{\textbf{Constraints from $\rho+p_r>0$}}\label{subsubsection1}
Here, we shall begin with the analysis of a combination $\rho+p_r>0$. Using Eq. \eqref{20}, the inequality $\rho+p_r>0$ with the restriction given in Eq. \eqref{27} can be written in the form
\begin{equation}\label{29}
\left(\frac{r_0}{r}\right)^{\eta +1 } \left(\eta -2 \lambda  \phi _0 \left(\frac{r_0}{r}\right)^{\lambda }+1\right)+2 \lambda  \phi _0 \left(\frac{r_0}{r}\right)^{\lambda }>0\,.
\end{equation}
By rearranging the above Eq. \eqref{29}, the parameter $\phi_0$ can be written in the combination of $\lambda,\,\eta,\,r,\,\text{and}  \,r_0$ as 
\begin{equation}\label{30}
\phi_0 > \frac{1+\eta}{2\lambda} \,\text{max}\left( \frac{\left(\frac{r_0}{r}\right)^{\eta +1 }}{\left(\frac{r_0}{r}\right)^{\lambda}\left( \left(\frac{r_0}{r}\right)^{\eta +1 }-1\right)}\right)  \equiv \phi_{\text{min}} \,. 
\end{equation}
At the throat $r=r_0$, the combination $\rho+p_r$ is positive if $\phi_0>\phi_c$ from Eq. \eqref{28}. Also, from the above Eq. \eqref{30}, one can see that $\phi_0$ is positive for the whole range of $r$ i.e. condition $\rho+p_r$ does not have any zeroes or does not change the sign if $\phi_0>\phi_{\text{min}}$. Thus, $\rho+p_r$ is always positive in the entire spacetime if $\phi_0> \text{max} (\phi_c,\, \phi_{\text{min}})$.

\subsubsection{\textbf{Constraints from $\rho+p_t>0$}}\label{subsubsection2}
Now, We shall have a look into the combination $\rho+p_t>0$. Using Eq. \eqref{21}, the inequality $\rho+p_t>0$ with the restriction given in Eq. \eqref{27} can be written in the form
\begin{equation}\label{31}
\left(\frac{r_0}{r}\right)^{\eta +1} \left(-\eta -\lambda  \phi _0 \left(\frac{r_0}{r}\right)^{\lambda } \left(\eta +2 \lambda +2 \lambda  \phi _0 \left(\frac{r_0}{r}\right)^{\lambda }+1\right)+1\right)+2 \lambda ^2 \phi _0 \left(\frac{r_0}{r}\right)^{\lambda } \left(\phi _0 \left(\frac{r_0}{r}\right)^{\lambda }+1\right) <0 \,.
\end{equation}
By rearranging the above Eq. \eqref{31} same as Eq. \eqref{29}, the parameter $\phi_0$ has some bound. However, Eq. \eqref{31} is quadratic in $\phi_0$, this equation imposes a double constraint on the value of $\phi_0$. Thus, the range of the parameter $\phi_0$ is given by 
\begin{equation}\label{32}
    \text{max}\left[h_-\left(\frac{r_0}{r}\right)\right]\,<\phi_0\,<\text{min}\left[h_+\left(\frac{r_0}{r}\right)\right] \,,
\end{equation}
where the functions $h_-\left(\frac{r_0}{r}\right)$ and $h_+\left(\frac{r_0}{r}\right)$ are given by
\begin{equation}\label{33}
    h_-\left(\frac{r_0}{r}\right)=\frac{-B_1\left(\frac{r_0}{r}\right)-\sqrt{B_1\left(\frac{r_0}{r}\right)^2-4A_1\left(\frac{r_0}{r}\right)C_1\left(\frac{r_0}{r}\right)}}{2A_1\left(\frac{r_0}{r}\right)} \,,
\end{equation}
\begin{equation}\label{34}
    h_+\left(\frac{r_0}{r}\right)=\frac{-B_1\left(\frac{r_0}{r}\right)+\sqrt{B_1\left(\frac{r_0}{r}\right)^2-4A_1\left(\frac{r_0}{r}\right)C_1\left(\frac{r_0}{r}\right)}}{2A_1\left(\frac{r_0}{r}\right)}\,,
\end{equation}
and the functions $A_1\left(\frac{r_0}{r}\right),\,B_1\left(\frac{r_0}{r}\right),\,\text{and}\,C_1\left(\frac{r_0}{r}\right)$ in the form of $\lambda\,\text{and}\,\eta$ are given by
\begin{equation}\label{35}
    A_1\left(\frac{r_0}{r}\right)= 2 \lambda ^2 \left(1-\left(\frac{r_0}{r}\right)^{\eta +1}\right) \left(\frac{r_0}{r}\right)^{2 \lambda } \,,
\end{equation}
\begin{equation}\label{36}
     B_1\left(\frac{r_0}{r}\right)=\lambda  \left(\frac{r_0}{r}\right)^{\lambda } \left(2 \lambda -(\eta +2 \lambda +1) \left(\frac{r_0}{r}\right)^{\eta +1}\right) \,, 
\end{equation}
\begin{equation}\label{37}
    C_1\left(\frac{r_0}{r}\right)= (1- \eta) \left(\frac{r_0}{r}\right)^{\eta +1} \,.
\end{equation}

At the throat $r=r_0$, the combination $\rho+p_t$ is positive if $\phi_0>\phi_c$ from Eq. \eqref{28}. Also, from the above Eq. \eqref{32}, one can see that $\phi_0$ is positive for the whole range of $r$ i.e. condition $\rho+p_t$ does not have any zeroes or does not change the sign if Eq. \eqref{32} holds. Thus, $\rho+p_t$ is always positive in the entire spacetime. One could see the function $h_-\left(\frac{r_0}{r}\right)$ is monotonically increasing in the interval $\frac{r_0}{r} \in (0, 1]$, which shows that $\text{max}\left[ h_-\left(\frac{r_0}{r}\right)\right] = h_-(1) = \phi_c$. Interestingly, while analyzing $\rho + p_t >0$, one could find that if ${B_1}^2- 4A_1 C_1= 0$ for some $\frac{r_0}{r}$ then $\text{max}\left[ h_-\left(\frac{r_0}{r}\right)\right] = \text{min}\left[ h_+\left(\frac{r_0}{r}\right)\right]$, but it gives a contradiction for Eq. \eqref{32} and is impossible to satisfy it and so there would not exist $\phi_0$ such that $\rho+p_t$ doesn't change sign.  

\subsubsection{\textbf{Full set of constraints and solution}}
 In the sections \ref{subsubsection1} and \ref{subsubsection2}, we established the conditions necessary for maintaining the combinations $\rho+p_r>0$ and $\rho+p_t> 0$ throughout space-time. By combining the results of our analysis, we have created a set of constraints on the parameters $\lambda,\,\eta,\,\text{and}\, \phi_0$ that enable us to find wormhole solutions that satisfy the NEC for the entire spacetime. This is in line with our initial assumption that either $\alpha <0,\,\, \beta >-8 \pi \,\,\,\,\text{or} \,\,\,\, \alpha >0,\,\, \beta <-8 \pi$, which we had previously established as necessary for the NEC to be valid at the throat. From the Sec. \ref{subsubsection1}, a necessary condition for $\rho+p_r>0$ is $\phi_0> \text{max} (\phi_c,\, \phi_{\text{min}})$, however from Sec. \ref{subsubsection2}, a necessary condition for $\rho+p_t>0$ is $\phi_c\,<\phi_0\,<\text{min}\left[h_+\left(\frac{r_0}{r}\right)\right]$. Furthermore, a necessary condition for satisfying both the combinations simultaneously is $\phi_c\,<\phi_0\,<\text{min}\left[h_+\left(\frac{r_0}{r}\right)\right]$ with $\text{min}\left[h_+\left(\frac{r_0}{r}\right)\right]> 
 \phi_{\text{min}}$. \\
 
 From equations \ref{30} and \ref{34}, it can be shown that $\phi_{\text{min}}=\text{min}\left[h_+\left(\frac{r_0}{r}\right)\right]=0$ in the parameter range $\lambda < \eta +1 $. In this range, $\phi_0 = 0$ is the only possible value for $\phi_0$ to allow the solutions satisfying NEC for whole spacetime, leading to the trivial redshift function $\phi (r) = 0$. Consequently, when $\lambda \geq \eta +1 $, the NEC is satisfied for the whole spacetime if $\phi_0 > 0$. This indicates that the redshift function should be non-trivial (i.e. not equal to zero) for physically relevant solutions to exist. It can be seen that this requirement of a non-trivial redshift function is consistent with the condition $\phi_c > \phi_{\text{min}}$, which must be fulfilled at all times. Therefore, we will restrict our analysis only to the parameter region in which $\phi_0 > 0$ in order to ensure the existence of physically meaningful solutions with non-trivial redshift functions. In particular, as highlighted in our earlier discussion, we need to consider parameters $\lambda, \eta,\, \text{and}\, \phi_0$ such that 
 \begin{align}\label{38}
     \phi_c\,&<\phi_0\,<\text{min}\left[h_+\left(\frac{r_0}{r}\right)\right],\,\,\,\eta>1,\,\,\,\lambda \geq \eta +1\,\,\,\,\,\,\,\,\,\,\,\,\,\,\text{with}\nonumber \\
     & (\alpha >0,\,\, \beta <-8 \pi)\,\,\,\,\text{or} \,\,\,\, (\alpha <0,\,\, \beta >-8 \pi) \,,
 \end{align}
 for wormhole solutions that satisfy the NEC for the entirety of the spacetime.\\

Subsequently, this analysis can be extended to encompass the verification of the Weak Energy Condition (WEC) and Strong Energy Condition (SEC) throughout the whole spacetime. Since these conditions are implied by NEC, a comprehensive analysis of the entire parameter space is unnecessary and we can instead limit our assessment to the already specified parameter region in Eq. \ref{38}. Therefore, in the following sections, we will focus on this particular parameter region to conduct a further investigation into the required constraints.

 \subsection{Conditions for WEC}\label{subsec2}
Let's turn to the WEC analysis. For WEC, the combinations  $\rho>0,\,\rho+p_r>0,\,\text{and}\,\rho+p_t>0$ must be satisfied. We have already discussed $\rho+p_r>0\,\text{and}\,\rho+p_t>0$ in the previous subsection \ref{subsec1}. So, in this subsection, we are going to study the positivity of energy density i.e. $\rho>0$. The following boundary condition applies at the throat
\begin{equation}\label{39}
    \rho(r) \bigg\vert_{(r=\,r_0)} = -\frac{\alpha  \left(\lambda  \phi _0 (-5 \beta  \eta +\beta +48 \pi )-4 \beta  \eta +48 \pi  \eta \right)}{12 (4 \pi -\beta ) (\beta +8 \pi ) {r_0}^2} \,.
\end{equation}
Here, one can clearly see that $\rho$ is always positive at $r=r_0$ under the conditions given in Eq. \eqref{38} except $\alpha <0,\,\, \beta >-8 \pi$. So, we are again restricting Eq. \eqref{38} to satisfy NEC in the whole spacetime and make $\rho$ positive at the throat into 
\begin{align}\label{40}
     \phi_c\,&<\phi_0\,<\text{min}\left[h_+\left(\frac{r_0}{r}\right)\right],\,\,\,\,\,\,\eta>1,\,\,\,\,\,\,\lambda \geq \eta +1\,\,\,\,\,\,\text{with}\,\,\,\,\,\,
     (\alpha >0,\,\, \beta <-8 \pi) \,.
 \end{align}
 In addition, $\rho>0$ along with Eq. \eqref{39}, impose  a constraint on parameter $\phi_0$ i.e.
\begin{equation}\label{41}
\phi _0>\frac{4\eta \left(\beta-12\pi\right) }{\lambda \left(48\pi+\beta-5\beta\eta\right)}\equiv \phi_{2c} \,.
\end{equation}
 Now, using Eq. \eqref{17}, the inequality $\rho>0$ with the restriction given in \eqref{40} can be written in the form
\begin{align}\label{42}
\left(\frac{r_0}{r}\right)^{\eta +1} \left(4 (\beta -12 \pi ) \eta +\lambda  \phi _0 \left(\frac{r_0}{r}\right)^{\lambda } \left(\beta  (5 \eta +10 \lambda -11)+10 \beta  \lambda  \phi _0 \left(\frac{r_0}{r}\right)^{\lambda }-48 \pi \right)\right)\nonumber\\
    -10 \beta  \lambda \phi _0 \left(\frac{r_0}{r}\right)^{\lambda } \left(\lambda +\lambda  \phi _0 \left(\frac{r_0}{r}\right)^{\lambda }-1\right)<0 \,.
\end{align}
By rearranging the above Eq. \eqref{42} same as Eq. \eqref{31} i.e. analysis of $\rho + p_t$, the parameter $\phi_0$ has some bound. However, Eq. \eqref{42} is quadratic in $\phi_0$, this equation imposes a double constraint on the value of $\phi_0$. Thus, the range of the parameter $\phi_0$ is given by
\begin{equation}\label{43}
    \text{max}\left[g_-\left(\frac{r_0}{r}\right)\right]\,<\phi_0\,<\text{min}\left[g_+\left(\frac{r_0}{r}\right)\right] \,,
\end{equation}
where the functions $g_-\left(\frac{r_0}{r}\right)$ and $g_+\left(\frac{r_0}{r}\right)$ are given by
\begin{equation}\label{44}
    g_-\left(\frac{r_0}{r}\right)=\frac{-B_2\left(\frac{r_0}{r}\right)-\sqrt{B_2\left(\frac{r_0}{r}\right)^2-4A_2\left(\frac{r_0}{r}\right)C_2\left(\frac{r_0}{r}\right)}}{2A_2\left(\frac{r_0}{r}\right)} \,,
\end{equation}
\begin{equation}\label{45}
    g_+\left(\frac{r_0}{r}\right)=\frac{-B_2\left(\frac{r_0}{r}\right)+\sqrt{B_2\left(\frac{r_0}{r}\right)^2-4A_2\left(\frac{r_0}{r}\right)C_2\left(\frac{r_0}{r}\right)}}{2A_2\left(\frac{r_0}{r}\right)} \,,
\end{equation}
and the functions $A_2\left(\frac{r_0}{r}\right),\,B_2\left(\frac{r_0}{r}\right),\,\text{and}\,C_2\left(\frac{r_0}{r}\right)$ in the form of $\lambda,\,\eta,\,\text{and}\,\beta$ are given by
\begin{equation}\label{46}
    A_2\left(\frac{r_0}{r}\right)= -10 \beta  \lambda ^2 \left(1-\left(\frac{r_0}{r}\right)^{\eta +1}\right) \left(\frac{r_0}{r}\right)^{2 \lambda } \,,
\end{equation}
\begin{equation}\label{47}
     B_2\left(\frac{r_0}{r}\right)= \lambda  \left(\frac{r_0}{r}\right)^{\lambda } \left(\left(\frac{r_0}{r}\right)^{\eta +1} (\beta  (5 \eta +10 \lambda -11)-48 \pi )-10 \beta  (\lambda -1)\right) \,,
\end{equation}
\begin{equation}\label{48}
    C_2 \left(\frac{r_0}{r}\right)= 4 (\beta- 12 \pi) \eta  \left(\frac{r_0}{r}\right){}^{\eta +1} \,.
\end{equation}
At the throat $r=r_0$, the energy density $\rho$ is positive under the restriction given in Eq. \eqref{40}. Also, from the above Eq. \eqref{43}, one can see that $\phi_0$ is positive for the whole range of $r$ i.e. $\rho$ does not have any zeroes or does not change the sign if Eq. \eqref{43} holds. Thus, the energy density $\rho$ is always positive in the entire spacetime. One could see the function $g_-\left(\frac{r_0}{r}\right)$ is monotonically increasing in the interval $\frac{r_0}{r} \in (0, 1]$, which shows that $\text{max}\left[ g_-\left(\frac{r_0}{r}\right)\right] = g_-(1) \equiv \phi_{2c}$.
Here, we can verify that $\phi_{c}>\phi_{2c}$ and $\text{min}\left[h_+\left(\frac{r_0}{r}\right)\right]<\text{min}\left[g_+\left(\frac{r_0}{r}\right)\right]$ under the restrictions given in Eq. \eqref{40}, it gives that bounds on $\phi_0$ arising from NEC are stronger than ones arising from $\rho>0$ i.e. if we choose parameters from restrictions obtained from NEC to satisfy the NEC through entire space, this process will keep $\rho$ positive everywhere and thus the WEC will also be satisfied through entire space. But the converse of this statement is not true i.e. $\rho>0 \nrightarrow$ verification of NEC. This one sided result guarantees that one doesn't get $ \text{max}\left[g_-\left(\frac{r_0}{r}\right)\right]\,=\,\text{min}\left[g_+\left(\frac{r_0}{r}\right)\right]$.

 \subsection{Conditions for  SEC}\label{subsec3}
Let's go toward the analysis of SEC. For SEC, the combinations  $\rho+p_r>0,\,\rho+p_t>0,\,\text{and}\,\rho+p_r+2p_t>0$ must be satisfied. We have already discussed $\rho+p_r>0\,\text{and}\,\rho+p_t>0$ in the previous subsection \ref{subsec1}. So, in this subsection, we are going to study the positivity of $\rho+p_r+2p_t$. The following boundary condition applies at the throat
\begin{equation}\label{49}
    \rho(r)+p_r(r)+2p_t(r) \bigg\vert_{(r=\,r_0)} = \frac{\alpha  \left(\lambda  \phi _0 (\beta  (\eta +7)-24 \pi  (\eta -1))+8 \beta  \eta \right)}{6 (4 \pi -\beta ) (\beta +8 \pi ) {r_0}^2} \,.
\end{equation}
Here, one can see that $\rho+p_r+2p_t$ is always positive at $r=r_0$ under the conditions given in Eq. \eqref{40}. In addition, $\rho+p_r+2p_t>0$ along with Eq. \eqref{49}, impose  a constraint on parameter $\phi_0$ i.e.
\begin{equation}\label{50}
\phi _0>-\frac{8 \beta  \eta }{\beta  \eta  \lambda +7 \beta  \lambda -24 \pi  \eta  \lambda +24 \pi  \lambda }\equiv \phi_{3c}  \,.
\end{equation}
 Now, using Eq. \eqref{24}, the inequality $\rho+p_r+2p_t>0$ with the restriction given in \eqref{40} can be written in the form
\begin{multline}\label{51}
\left(\frac{r_0}{r}\right)^{\eta +1} \left(8 \beta  \eta +\lambda  \phi _0 \left(\frac{r_0}{r}\right)^{\lambda } \left(\beta  (\eta +2 \lambda +5)-24 \pi  (\eta +2 \lambda -3)-2 (24 \pi -\beta ) \lambda  \phi _0 \left(\frac{r_0}{r}\right)^{\lambda }\right)\right)\\
+2 (24 \pi -\beta ) \lambda  \phi _0 \left(\frac{r_0}{r}\right)^{\lambda } \left(\lambda +\lambda  \phi _0 \left(\frac{r_0}{r}\right)^{\lambda }-1\right)<0  \,.
\end{multline}
By rearranging the above Eq. \eqref{51} same as Eq. \eqref{31} i.e. analysis of $\rho + p_t$, the parameter $\phi_0$ has some bound. However, Eq. \eqref{51} is quadratic in $\phi_0$, this equation imposes a double constraint on the value of $\phi_0$. Thus, the range of the parameter $\phi_0$ is given by
\begin{equation}\label{52}
    \text{max}\left[f_-\left(\frac{r_0}{r}\right)\right]\,<\phi_0\,<\text{min}\left[f_+\left(\frac{r_0}{r}\right)\right] \,,
\end{equation}
where the functions $f_-\left(\frac{r_0}{r}\right)$ and $f_+\left(\frac{r_0}{r}\right)$ are given by
\begin{equation}\label{53}
    f_-\left(\frac{r_0}{r}\right)=\frac{-B_3\left(\frac{r_0}{r}\right)-\sqrt{B_3\left(\frac{r_0}{r}\right)^2-4A_3\left(\frac{r_0}{r}\right)C_3\left(\frac{r_0}{r}\right)}}{2A_3\left(\frac{r_0}{r}\right)} \,,
\end{equation}
\begin{equation}\label{54}
    f_+\left(\frac{r_0}{r}\right)=\frac{-B_3\left(\frac{r_0}{r}\right)+\sqrt{B_3\left(\frac{r_0}{r}\right)^2-4A_3\left(\frac{r_0}{r}\right)C_3\left(\frac{r_0}{r}\right)}}{2A_3\left(\frac{r_0}{r}\right)} \,,
\end{equation}
and the functions $A_3\left(\frac{r_0}{r}\right),\,B_3\left(\frac{r_0}{r}\right),\,\text{and}\,C_3\left(\frac{r_0}{r}\right)$ in the form of $\lambda,\,\eta,\,\text{and}\,\beta$ are given by
\begin{equation}\label{55}
    A_3\left(\frac{r_0}{r}\right)= 2 (24 \pi -\beta ) \lambda ^2 \left(1-\left(\frac{r_0}{r}\right)^{\eta +1}\right) \left(\frac{r_0}{r}\right)^{2 \lambda } \,,
\end{equation}
\begin{equation}\label{56}
     B_3\left(\frac{r_0}{r}\right)=\lambda  \left(\frac{r_0}{r}\right)^{\lambda } \left(2 (24 \pi -\beta ) (\lambda -1) -\left(\frac{r_0}{r}\right)^{\eta +1} (24 \pi  (\eta +2 \lambda -3)-\beta  (\eta +2 \lambda +5))\right) \,,
\end{equation}
\begin{equation}\label{57}
    C_3 \left(\frac{r_0}{r}\right)= 8 \beta  \eta  
 \left(\frac{r_0}{r}\right)^{\eta +1} \,.
\end{equation}
At the throat $r=r_0$, the $\rho+p_r+2p_t$ is positive under the restriction given in Eq. \eqref{40}. Also, from the above Eq. \eqref{52}, one can see that $\rho+p_r+2p_t$ is positive for the whole range of $r$ i.e. $\rho+p_r+2p_t$ does not have any zeroes or does not change the sign if Eq. \eqref{52} holds. Thus, $\rho+p_r+2p_t$ is always positive in the entire spacetime. One could see the function $f_-\left(\frac{r_0}{r}\right)$ is monotonically increasing in the interval $\frac{r_0}{r} \in (0, 1]$, which shows that $\text{max}\left[ f_-\left(\frac{r_0}{r}\right)\right] = f_-(1) \equiv \phi_{3c}$.
Here, we can verify that $\phi_{c}>\phi_{4c}$ and $\text{min}\left[h_+\left(\frac{r_0}{r}\right)\right]<\text{min}\left[f_+\left(\frac{r_0}{r}\right)\right]$ under the restrictions given in Eq. \eqref{40}, it gives that bounds on $\phi_0$ arising from NEC are stronger than ones arising from $\rho+p_r+2p_t >0$ i.e. if we choose parameters from restrictions obtained from NEC to satisfy the NEC through entire space, this process will keep $\rho+p_r+2p_t$ positive everywhere and thus the SEC will also be satisfied through entire space. But the converse of this statement is not true i.e. $\rho+p_r+2p_t >0 \nrightarrow$ verification of NEC. This one sided result guarantees that one doesn't get $ \text{max}\left[f_-\left(\frac{r_0}{r}\right)\right]\,=\,\text{min}\left[f_+\left(\frac{r_0}{r}\right)\right]$.

 \subsection{Conditions for DEC}\label{subsec4}
 Finally, let's analyze the SEC. For SEC, the combinations  $\rho>0,\,\rho+p_r>0,\,\rho+p_t>0,\,\rho-p_r>0,\,\text{and}\,\rho-p_t>0$ must be satisfied. We have already discussed $\rho>0,\,\rho+p_r>0,\,\text{and}\,\rho+p_t>0$ in the previous subsections \ref{subsec1}-\ref{subsec2}. So, in this subsection, we are going to study the positivity of $\rho-p_r\,\text{and}\,\rho-p_t$. The following boundary condition applies at the throat for $\rho-p_r$
 \begin{equation}\label{58}
   \rho(r)-p_r(r) \bigg\vert_{(r=\,r_0)} = -\frac{\alpha  \left(\lambda  \phi _0 (-5 \beta  \eta +\beta +48 \pi )+2 \beta  (\eta +3)+24 \pi  (\eta -1)\right)}{6 (4 \pi -\beta ) (\beta +8 \pi ) r_0^2} \,,
 \end{equation}
\begin{equation}\label{59}
    \rho(r)-p_t(r) \bigg\vert_{(r=\,r_0)} = \frac{\alpha  \left(2 \lambda  \phi _0 (\beta  (\eta -2)+6 \pi  (\eta -3))+\beta  (\eta +3)-12 \pi  (3 \eta +1)\right)}{6 (4 \pi -\beta ) (\beta +8 \pi ) {r_0}^2} \,.
\end{equation}
Here, one can see that $\rho-p_t$ is always positive at $r=r_0$ under the conditions given in Eq. \eqref{40} and doesn't need any extra conditions other than those given in Eq. \eqref{40}. On the other hand, $\rho-p_r$ along with Eq. \eqref{58} impose  a constraint on parameter $\phi_0$ to be positive at the throat i.e.
\begin{equation}\label{60}
\phi _0>\frac{2 \beta  \eta +6 \beta +24 \pi  \eta -24 \pi }{5 \beta  \eta  \lambda -\beta  \lambda -48 \pi  \lambda }\equiv \phi_{4c}  \,.
\end{equation}
By taking the combination $\rho-p_r$ at the throat $r=r_0$ i.e. Eq. \eqref{58} along with the restrictions given in Eqs. \eqref{40} and \eqref{60}, one can verify the positivity of the combination $\rho - p_r$ at the throat. To guarantee the physical relevance of the obtained wormhole solutions, it is not enough to satisfy the DEC at the throat. Thus, to guarantee the physical relevance of the obtained wormhole solutions for the whole spacetime, we will start analysis for the combination $\rho-p_r$ and $\rho-p_t$ separately and impose constraints on the parameters $\beta$, $\lambda$, $\eta$, and $\phi_0$ which will guarantee for $\rho-p_r>0$ and $\rho-p_t>0$. Then we will combine the results into a unified set of constraints.

\subsubsection{\textbf{Constraints from $\rho-p_r>0$}}
Here, we shall begin with the analysis of a combination $\rho-p_r>0$. Using Eq. \eqref{22}, the inequality $\rho-p_r>0$ with the restriction given in Eq. \eqref{40} can be written in the form
\begin{multline}\label{62}
\left(\frac{r_0}{r}\right)^{\eta +1} \left(-2 \beta  (\eta +3)-24 \pi  (\eta -1)+\lambda  \phi _0 \left(\frac{r_0}{r}\right)^{\lambda } \left(5 \beta  \eta +10 \beta  \lambda +\beta +10 \beta  \lambda  \phi _0 \left(\frac{r_0}{r}\right)^{\lambda }-96 \pi \right)\right)\\
-2 \lambda  \phi _0 \left(\frac{r_0}{r}\right)^{\lambda } \left(5 \beta  \lambda +\beta +5 \beta  \lambda  \phi _0 \left(\frac{r_0}{r}\right)^{\lambda }-24 \pi \right)<0  \,.
\end{multline}
Same as the previous one, above Eq. \eqref{62} is quadratic in $\phi_0$ and so imposes a double constraint on the value of $\phi_0$. Thus, the range of the parameter $\phi_0$ is given by
\begin{equation}\label{63}
    \text{max}\left[F_-\left(\frac{r_0}{r}\right)\right]\,<\phi_0\,<\text{min}\left[F_+\left(\frac{r_0}{r}\right)\right] \,,
\end{equation}
where the functions $F_-\left(\frac{r_0}{r}\right)$ and $F_+\left(\frac{r_0}{r}\right)$ are given by
\begin{equation}\label{64}
    F_-\left(\frac{r_0}{r}\right)=\frac{-B_4\left(\frac{r_0}{r}\right)-\sqrt{B_4\left(\frac{r_0}{r}\right)^2-4A_4\left(\frac{r_0}{r}\right)C_4\left(\frac{r_0}{r}\right)}}{2A_4\left(\frac{r_0}{r}\right)} \,,
\end{equation}
\begin{equation}\label{65}
    F_+\left(\frac{r_0}{r}\right)=\frac{-B_4\left(\frac{r_0}{r}\right)+\sqrt{B_4\left(\frac{r_0}{r}\right)^2-4A_4\left(\frac{r_0}{r}\right)C_4\left(\frac{r_0}{r}\right)}}{2A_4\left(\frac{r_0}{r}\right)} \,,
\end{equation}
and the functions $A_4\left(\frac{r_0}{r}\right),\,B_4\left(\frac{r_0}{r}\right),\,\text{and}\,C_4\left(\frac{r_0}{r}\right)$ in the form of $\lambda,\,\eta,\,\text{and}\,\beta$ are given by
\begin{equation}\label{66}
    A_4\left(\frac{r_0}{r}\right)= -10 \beta  \lambda ^2 \left(1-\left(\frac{r_0}{r}\right)^{\eta +1}\right) \left(\frac{r_0}{r}\right)^{2 \lambda } \,,
\end{equation}
\begin{equation}\label{67}
     B_4\left(\frac{r_0}{r}\right)=\lambda  \left(\frac{r_0}{r}\right)^{\lambda } \left(\left(\frac{r_0}{r}\right)^{\eta +1} (5 \beta  \eta +10 \beta  \lambda +\beta -96 \pi )-2 (5 \beta  \lambda +\beta -24 \pi )\right) \,,
\end{equation}
\begin{equation}\label{68}
    C_4 \left(\frac{r_0}{r}\right)= -2 (\beta  (\eta +3)+12 \pi  (\eta -1)) \left(\frac{r_0}{r}\right){}^{\eta +1} \,.
\end{equation}
Thus, the $\rho-p_r$ is positive at the throat $r=r_0$ under the restriction given in Eqs. \eqref{40} with the extra restriction on $\phi_0$ in Eq. \eqref{60}. Also, from the above Eq. \eqref{63}, one can see that $\rho-p_r$ is positive for the whole range of $r$ i.e. $\rho-p_r$ does not have any zeroes or does not change the sign if Eq. \eqref{63} holds. Thus, $\rho-p_r$ is always positive in the entire spacetime. One could see the function $F_-\left(\frac{r_0}{r}\right)$ is monotonically increasing in the interval $\frac{r_0}{r} \in (0, 1]$, which shows that $\text{max}\left[ F_-\left(\frac{r_0}{r}\right)\right] = F_-(1) \equiv \phi_{4c}$.
Additionally, in this case, we need to verify if one gets $B_4\left(\frac{r_0}{r}\right)^2-4A_4\left(\frac{r_0}{r}\right)C_4\left(\frac{r_0}{r}\right)=0$ at any point $\frac{r_0}{r}$, which corresponds to $\text{max}\left[F_-\left(\frac{r_0}{r}\right)\right]\,=\,\text{min}\left[F_+\left(\frac{r_0}{r}\right)\right]$ and prevents from getting a suitable value of $\phi_0$. Now, taking a coordinate transformation as $\left(\frac{r_0}{r}\right)^{1+\eta}=x$, one can rewrite the equation $B_4\left(\frac{r_0}{r}\right)^2-4A_4\left(\frac{r_0}{r}\right)C_4\left(\frac{r_0}{r}\right)=0$ in the form 
\begin{equation}\label{69}
\left[-10 \beta  \lambda -2 \beta +x (5 \beta  \eta +10 \beta  \lambda +\beta -96 \pi )+48 \pi \right]^2+80 \beta  \left[\beta  (\eta +3)+12 \pi  (\eta -1)\right] \left(x-1\right) x=0 \,.
\end{equation}
Above Eq. \eqref{69} is quadratic in $x$, thus it gives two roots $x_1$ and $x_2$ and one can verify that these roots $x_1$ and $x_2$ are real and belongs to the interval $(0,1]$ under the restrictions given in Eq. \eqref{40} which corresponds to $\text{max}\left[F_-\left(\frac{r_0}{r}\right)\right]\,=\,\text{min}\left[F_+\left(\frac{r_0}{r}\right)\right]$. So one must need extra restrictions other than restrictions given in Eq. \eqref{40} to avoid these roots, one impose a constraint on $\beta$ and $\eta$ in the form 
\begin{equation}\label{70}
\frac{12 \pi -12 \pi  \eta }{\eta +3}<\beta <-8 \pi  \,\,\,\,\,\,\text{and}\,\,\,\,\,\,\eta>9 \,.
\end{equation} 
Now, by using the above extra restrictions given in Eq. \eqref{70} with Eq. \eqref{40} and \eqref{60}, one guarantees that $\text{max}\left[F_-\left(\frac{r_0}{r}\right)\right]\,=\,\text{min}\left[F_+\left(\frac{r_0}{r}\right)\right]$ doesn't occur and as $\text{max}\left[F_-\left(\frac{r_0}{r}\right)\right]=\phi_{4c}$, one guarantees the positivity of $\rho- p_r$ for the entire spacetime.

\subsubsection{\textbf{Constraints from $\rho-p_t>0$}}
Now, we shall have a look into the combination $\rho-p_t>0$. Using Eq. \eqref{23}, the inequality $\rho-p_r>0$ with the restriction given in Eq. \eqref{40} can be written in the form
\begin{multline}\label{71}
\left(\frac{r_0}{r}\right)^{\eta +1} \left(\beta  (\eta +3)-12 \pi  (3 \eta +1)+2 \lambda  \phi _0 \left(\frac{r_0}{r}\right)^{\lambda } \left(\beta  (\eta +2 \lambda -7)+6 \pi  (\eta +2 \lambda -3)+2 (\beta +6 \pi ) \lambda  \phi _0 \left(\frac{r_0}{r}\right)^{\lambda }\right)\right)\\
+2 \lambda  \phi _0 \left(\frac{r_0}{r}\right)^{\lambda } \left(-2 (\beta +6 \pi ) \lambda +5 \beta -2 (\beta +6 \pi ) \lambda  \phi _0 \left(\frac{r_0}{r}\right)^{\lambda }\right)<0 \,.
\end{multline}
By rearranging the above Eq. \eqref{71} same as Eq. \eqref{31} i.e. analysis of $\rho + p_t$, the parameter $\phi_0$ has some bound. However, Eq. \eqref{71} is quadratic in $\phi_0$, this equation imposes a double constraint on the value of $\phi_0$. Thus, the range of the parameter $\phi_0$ is given by
\begin{equation}\label{72}
    \text{max}\left[G_-\left(\frac{r_0}{r}\right)\right]\,<\phi_0\,<\text{min}\left[G_+\left(\frac{r_0}{r}\right)\right] \,,
\end{equation}
where the functions $G_-\left(\frac{r_0}{r}\right)$ and $G_+\left(\frac{r_0}{r}\right)$ are given by
\begin{equation}\label{73}
    G_-\left(\frac{r_0}{r}\right)=\frac{-B_5\left(\frac{r_0}{r}\right)-\sqrt{B_5\left(\frac{r_0}{r}\right)^2-4A_5\left(\frac{r_0}{r}\right)C_5\left(\frac{r_0}{r}\right)}}{2A_5\left(\frac{r_0}{r}\right)} \,, \,,
\end{equation}
\begin{equation}\label{74}
    G_+\left(\frac{r_0}{r}\right)=\frac{-B_5\left(\frac{r_0}{r}\right)+\sqrt{B_5\left(\frac{r_0}{r}\right)^2-4A_5\left(\frac{r_0}{r}\right)C_5\left(\frac{r_0}{r}\right)}}{2A_5\left(\frac{r_0}{r}\right)} \,,
\end{equation}
and the functions $A_5\left(\frac{r_0}{r}\right),\,B_5\left(\frac{r_0}{r}\right),\,\text{and}\,C_5\left(\frac{r_0}{r}\right)$ in the form of $\lambda,\,\eta,\,\text{and}\,\beta$ are given by
\begin{equation}\label{75}
    A_5\left(\frac{r_0}{r}\right)= 4 (\beta +6 \pi ) \lambda ^2 \left(\left(\frac{r_0}{r}\right)^{\eta +1}-1\right) \left(\frac{r_0}{r}\right)^{2 \lambda } \,,
\end{equation}
\begin{equation}\label{76}
     B_5\left(\frac{r_0}{r}\right)=2 \lambda  \left(\frac{r_0}{r}\right)^{\lambda } \left(\left(\frac{r_0}{r}\right)^{\eta +1} (\beta  (\eta +2 \lambda -7)+6 \pi  (\eta +2 \lambda -3))-2 (\beta +6 \pi ) \lambda +5 \beta \right) \,,
\end{equation}
\begin{equation}\label{77}
    C_5 \left(\frac{r_0}{r}\right)= (\beta  (\eta +3)-12 \pi  (3 \eta +1)) \left(\frac{r_0}{r}\right)^{\eta +1} \,.
\end{equation}
Here, one verifies that roots of equation $B_5\left(\frac{r_0}{r}\right)^2-4A_5\left(\frac{r_0}{r}\right)C_5\left(\frac{r_0}{r}\right)=0$ are not lie in the interval $(0,1]$ and also doesn't get $\text{max}\left[G_-\left(\frac{r_0}{r}\right)\right]\,=\,\text{min}\left[G_+\left(\frac{r_0}{r}\right)\right]$. Thus, $\rho-p_t$ is positive in the entire spacetime with the restrictions as
\begin{equation}\label{78}
\phi_0\,>\,\phi_{4c},\,\,\,\,\eta>9,\,\,\,\,\lambda \geq \eta+1, \,\,\,\, \text{and} \,\,\,\,\frac{12 \pi -12 \pi  \eta }{\eta +3}<\beta <-8 \pi  \,.
\end{equation}
Hence, one can see that DEC requires more restrictions than NEC to be satisfied throughout the whole spacetime. Now, in the next subsection, we will combine all the necessary conditions to satisfy the NEC, WEC, SEC, and DEC. 

\subsection{Explicit examples of solutions}\label{subsec5}
\begin{figure}[h]
\centering
\includegraphics[width=12.5cm,height=8cm]{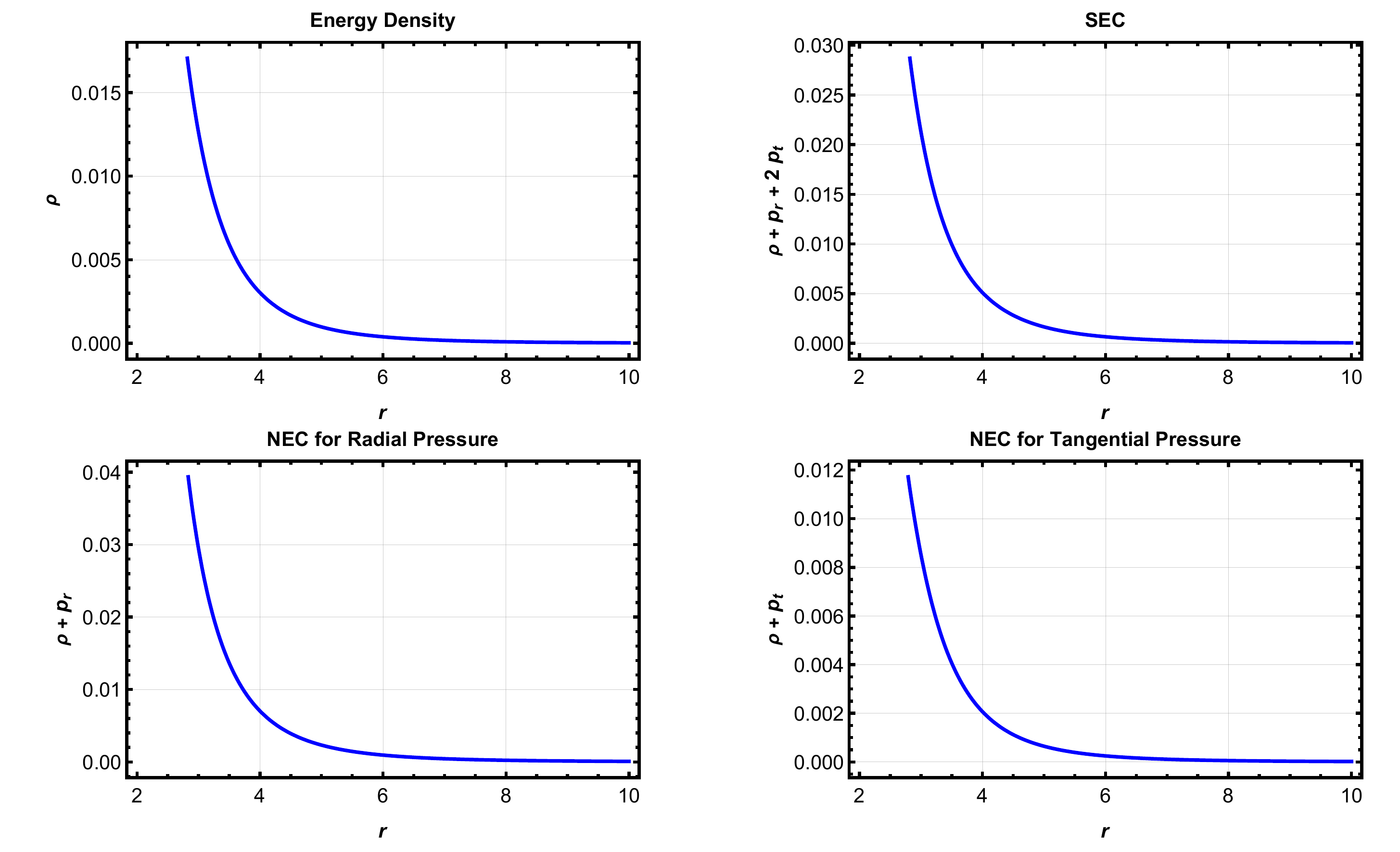}
\caption{Profile shows the behavior of NEC, WEC, and SEC for $\alpha =1$, $\beta =-9 \pi$, $\eta =2$, $\lambda =4$, $r_0=2$, and $\phi_0=-0.05$\,. }
\label{fig1}
\end{figure}

\begin{figure}[h]
\centering
\includegraphics[width=15.5cm,height=8cm]{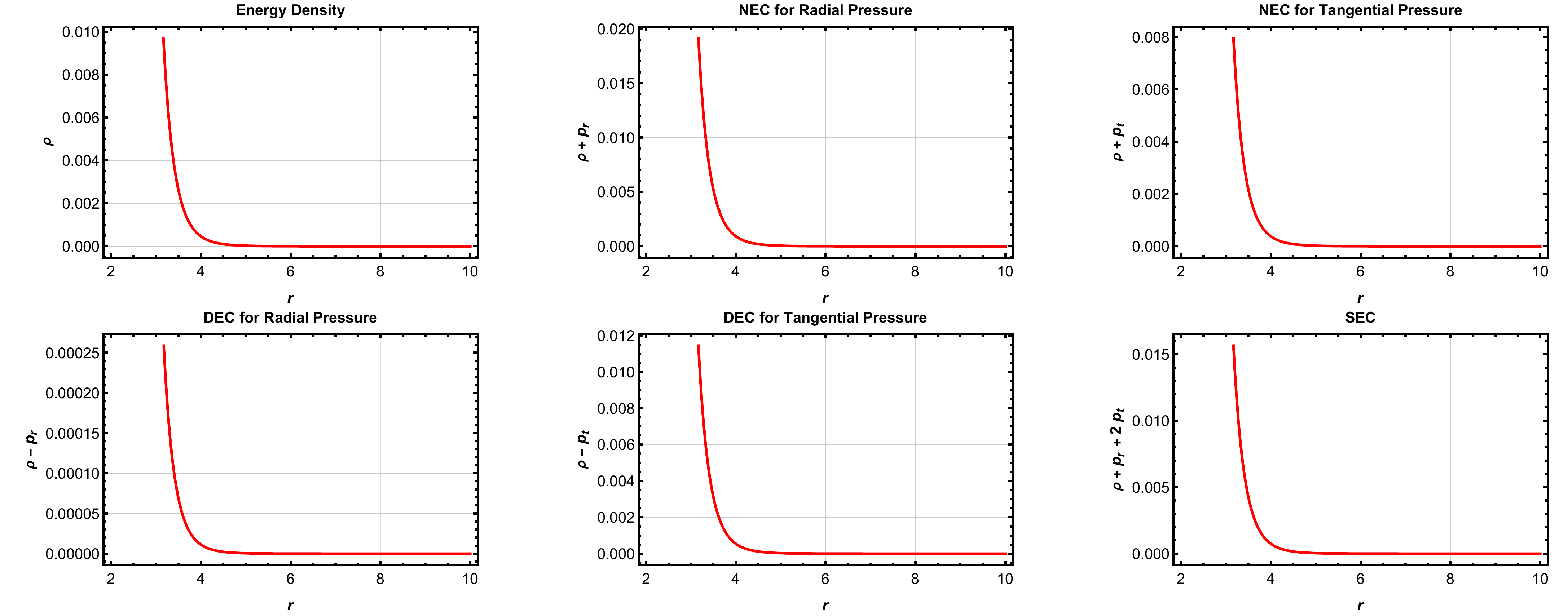}
\caption{Profile shows the behavior of NEC, WEC, SEC, and DEC for $\alpha =1$, $\beta =-25.5$, $\eta =10$, $\lambda =12$, $r_0=2$, and $\phi_0=-0.0008$\,.}
\label{fig2}
\end{figure}

In the preceding subsections \ref{subsec1}-\ref{subsec4}, we have analyzed the wormhole solution and formed some necessary conditions for the energy conditions to be satisfied. They are summarised as follows:\\
\textit{Solution satisfying NEC:}
\begin{minipage}[t]{10cm}
\begin{enumerate}
\item Choose $(\alpha >0,\,\, \beta <-8 \pi)\,\,\,\,\text{or} \,\,\,\, (\alpha <0,\,\, \beta >-8 \pi)$ \,,
\item Choose $\eta>1$ and $\lambda \geq \eta +1$ \,,
\item Choose $\phi_c\,<\phi_0\,<\text{min}\left[h_+\left(\frac{r_0}{r}\right)\right]$ \,.
\end{enumerate}
\end{minipage}\\
\textit{Solution satisfying NEC and WEC:}
\begin{minipage}[t]{10cm}
\begin{enumerate}
\item Choose $\alpha >0$ and $\beta <-8 \pi$ \,,
\item Choose $\eta>1$ and $\lambda \geq \eta +1$ \,,
\item Choose $\phi_c\,<\phi_0\,<\text{min}\left[h_+\left(\frac{r_0}{r}\right)\right]$ \,.
\end{enumerate}
\end{minipage}\\
\textit{Solution satisfying NEC, WEC, and SEC:}
\begin{minipage}[t]{10cm}
\begin{enumerate}
\item Choose $\alpha >0$ and $\beta <-8 \pi$ \,,
\item Choose $\eta>1$ and $\lambda \geq \eta +1$ \,,
\item Choose $\phi_c\,<\phi_0\,<\text{min}\left[h_+\left(\frac{r_0}{r}\right)\right]$ \,.
\end{enumerate}
\end{minipage}\\
\textit{Solution satisfying NEC, WEC, SEC and DEC:}
\begin{minipage}[t]{10cm}
\begin{enumerate}
\item Choose $\alpha >0$ and $\beta <-8 \pi$ \,,
\item Choose $\eta>9$ \,\,such that\,\, $\frac{12 \pi -12 \pi  \eta }{\eta +3}<\beta <-8 \pi $ \,,
\item Choose $\lambda \geq \eta +1$ \,,
\item Choose $\phi_{4c}\,<\phi_0\,<\text{min}\left[F_+\left(\frac{r_0}{r}\right)\right]$ \,.
\end{enumerate}
\end{minipage}\\
Now, considering the above analysis, we shall explore two examples of solutions, out of that one satisfying NEC, WEC, and SEC in Fig. \ref{fig1}; and another is for the solution satisfying all the energy conditions i.e. NEC, WEC, SEC, and DEC in Fig. \ref{fig2} by considering some particular values of $\alpha$, $\beta$, $\lambda$, $\eta$, and $\phi_0$.

\section{Wormhole solutions with non-linear $f(Q,T)$}\label{sec5}
In this section, we consider non-linear form of $f(Q,T)$ as  \cite{Dixit}
\begin{equation}
\label{79}
f(Q,T)=Q+\gamma\,Q^2+\mu\,T \,,
\end{equation}
 where $\gamma$ and $\mu$ are model parameters.\\
 Also, we are choosing the same redshift function $\phi(r)$ and shape function $b(r)$ as used in the linear model given by Eqs. \eqref{15} and  \eqref{16}. Using a non-linear form of $f(Q,T)$, particular form of the redshift and the shape function; energy density $\rho$, radial pressure $p_r$, and tangential pressure $p_t$ are calculated from the field equations \eqref{11}-\eqref{13} and written as
\begin{multline}\label{80}
    \rho = \frac{1}{12 (4 \pi -\mu ) (\mu +8 \pi ) r^6 \left(r-r_0 \left(\frac{r_0}{r}\right)^{\eta }\right)^2}\left[r r_0^3 \left(\frac{r_0}{r}\right)^{3 \eta } \left(\lambda  \phi _0 \left(\frac{r_0}{r}\right)^{\lambda } \left(\mu  \left(2 \gamma  \left(5 \eta ^2+20 (\eta +1) \lambda +34 \eta +49\right) 
    \right.\right.\right.\right.\\ \left.\left.\left.\left.
     \hspace{0.5cm}+r^2 (5 \eta +10 \lambda -11)\right)+10 \lambda  \mu  \phi _0 \left(\frac{r_0}{r}\right)^{\lambda } \left(4 \gamma  (\eta +1)+r^2\right)-48 \pi  \left(2 \gamma  (\eta +1)+r^2\right)\right)-4 (12 \pi -\mu ) \left(2 \gamma  (\eta +2)^2
   \right.\right.\right. \\ \left.\left.\left.
    \hspace{0.5cm} +\eta  r^2\right)\right) +2 r^2 r_0^2 \left(\frac{r_0}{r}\right)^{2 \eta } \left(\lambda  \phi _0 \left(\frac{r_0}{r}\right)^{\lambda } \left(-5 \lambda  \mu  \phi _0 \left(\frac{r_0}{r}\right)^{\lambda } \left(2 \gamma  (\eta +1)+3 r^2\right)-\mu  \left(10 \gamma  (\eta  (\lambda +2)+\lambda +3)+r^2 (5 \eta
    \right.\right.\right.\right. \\ \left.\left.\left.\left.
   \hspace{0.5cm}   +15 \lambda -16)\right)+48 \pi  r^2\right) +4 \eta  (12 \pi -\mu ) r^2\right)+r^6 \left(\frac{r_0}{r}\right)^{\eta +1} \left(4 \eta  (\mu -12 \pi )+\lambda  \phi _0 \left(\frac{r_0}{r}\right)^{\lambda } \left(\mu  (5 \eta +30 \lambda -31)+30 \lambda  
     \right.\right.\right. \\ \left.\left.\left.
     \hspace{0.5cm}  \mu  \phi _0 \left(\frac{r_0}{r}\right)^{\lambda }-48 \pi \right)\right)-10 \lambda  \mu  r^6 \phi _0 \left(\frac{r_0}{r}\right)^{\lambda } \left(\lambda +\lambda  \phi _0 \left(\frac{r_0}{r}\right)^{\lambda }-1\right)+\gamma  r_0^4 \left(\frac{r_0}{r}\right)^{4 \eta } \left(-(\eta  (11 \eta +30)+27) \mu +24 \pi  
      \right.\right. \\ \left.\left.
     (\eta  (3 \eta +10)+11)+2 \lambda  \phi _0 \left(\frac{r_0}{r}\right)^{\lambda }\left(-\mu  \left(5 \eta ^2+10 (\eta +1) \lambda +14 \eta +19\right)+48 \pi  (\eta +1)-10 (\eta +1) \lambda  \mu  \phi _0 \left(\frac{r_0}{r}\right)^{\lambda }\right)\right)\right] \,,
\end{multline}
\begin{multline}\label{81}
    p_r = \frac{1}{12 (4 \pi -\mu ) (\mu +8 \pi ) r^6 \left(r-r_0 \left(\frac{r_0}{r}\right)^{\eta }\right)^2}\left[r r_0^3 \left(\frac{r_0}{r}\right)^{3 \eta } \left(\lambda  \phi _0 \left(\frac{r_0}{r}\right)^{\lambda } \left(-\mu  \left(2 \gamma  \left(5 \eta ^2+20 (\eta +1) \lambda +82 \eta +97\right) 
     \right.\right.\right.\right.\\ \left.\left.\left.\left.
    \hspace{0.4cm} +r^2 (5 \eta +10 \lambda +13)\right)-10 \lambda  \mu  \phi _0 \left(\frac{r_0}{r}\right)^{\lambda } \left(4 \gamma  (\eta +1)+r^2\right)+48 \pi  \left(10 \gamma  (\eta +1)+3 r^2\right)\right)+4 \mu  \left(4 \gamma  \eta  (\eta +1)-2 \gamma +
   \right.\right.\right.\\ \left.\left.\left.
    \hspace{0.5cm}(2 \eta +3) r^2\right)-48 \pi  \left(r^2-2 \gamma  (2 \eta +3)\right)\right)+2 r^2 r_0^2 \left(\frac{r_0}{r}\right)^{2 \eta } \left(\lambda  \phi _0 \left(\frac{r_0}{r}\right)^{\lambda } \left(2 \gamma  \mu  (5 (\eta +1) \lambda +22 \eta +27)+5 \lambda  \mu  \phi _0 \left(\frac{r_0}{r}\right)^{\lambda } 
    \right.\right.\right.\\ \left.\left.\left.
    \hspace{0.5cm}\left(2 \gamma  (\eta +1)+3 r^2\right)-96 \pi  \left(\gamma  \eta +\gamma +2 r^2\right)+5 \mu  r^2 (\eta +3 \lambda +4)\right)+4 r^2 (12 \pi -(2 \eta +3) \mu )\right)+r^6 \left(\frac{r_0}{r}\right)^{\eta +1} \left(4 (2 \eta +3)
    \right.\right.\\ \left.\left.
    \hspace{0.5cm} \mu -\lambda  \phi _0 \left(\frac{r_0}{r}\right)^{\lambda } \left(\mu  (5 \eta +30 \lambda +41)+30 \lambda  \mu  \phi _0 \left(\frac{r_0}{r}\right)^{\lambda }-336 \pi \right)-48 \pi \right)+2 \lambda  r^6 \phi _0 \left(\frac{r_0}{r}\right)^{\lambda } \left(5 \lambda  \mu +7 \mu +5 \lambda  \mu  \phi _0 \left(\frac{r_0}{r}\right)^{\lambda }
    \right.\right.\\ \left.\left.
    \hspace{0.5cm}-48 \pi \right)+\gamma  r_0^4 \left(\frac{r_0}{r}\right)^{4 \eta } \left((3-\eta  (13 \eta +18)) \mu +24 \pi  ((\eta -2) \eta -7)+2 \lambda  \phi _0 \left(\frac{r_0}{r}\right)^{\lambda } \left(\mu  \left(5 \eta ^2+10 (\eta +1) \lambda +38 \eta +43\right)
    \right.\right.\right.\\ \left.\left.\left.
   -144 \pi  (\eta +1)+10 (\eta +1) \lambda  \mu  \phi _0 \left(\frac{r_0}{r}\right)^{\lambda }\right)\right)\right] \,,
\end{multline}
\begin{multline}\label{82}
    p_t = \frac{1}{12 (4 \pi -\mu ) (\mu +8 \pi ) r^6 \left(r-r_0 \left(\frac{r_0}{r}\right)^{\eta }\right)^2}\left[2 r^2 r_0^2 \left(\frac{r_0}{r}\right)^{2 \eta } \left(\lambda  \phi _0 \left(\frac{r_0}{r}\right)^{\lambda } \left(\lambda  (24 \pi -\mu ) \phi _0 \left(\frac{r_0}{r}\right)^{\lambda } \left(2 \gamma  (\eta +1)+3 r^2\right)
     \right.\right.\right.\\ \left.\left.\left.
     \hspace{0.4cm} -\mu  \left(2 \gamma  (\eta  (\lambda +8)+\lambda +9)+r^2 (\eta +3 \lambda +22)\right)+24 \pi  \left(2 \gamma  (\eta  (\lambda +3)+\lambda +4)+r^2 (\eta +3 \lambda -1)\right)\right)-2 r^2 ((\eta -3) \mu
     \right.\right.\\ \left.\left.
      \hspace{0.4cm}+12 \pi  (\eta +1))\right)+r r_0^3 \left(\frac{r_0}{r}\right)^{3 \eta } \left(\lambda  \phi _0 \left(\frac{r_0}{r}\right)^{\lambda } \left(-2 \lambda  (24 \pi -\mu ) \phi _0 \left(\frac{r_0}{r}\right)^{\lambda } \left(4 \gamma  (\eta +1)+r^2\right)+\mu  \left(2 \gamma  (\eta  (\eta +4 \lambda +38)+
      \right.\right.\right.\right.\\ \left.\left.\left.\left.
     \hspace{0.4cm} 4 \lambda +41)+r^2 (\eta +2 \lambda +17)\right)-24 \pi  \left(2 \gamma  (\eta  (\eta +4 \lambda +10)+4 \lambda +13)+r^2 (\eta +2 \lambda -1)\right)\right)+2 \mu  \left(2 \gamma  (\eta  (\eta +10)+13)
     \right.\right.\right.\\ \left.\left.\left.
     \hspace{0.4cm}+(\eta -3) r^2\right)+24 \pi  (\eta +1) \left(2 \gamma  (\eta +1)+r^2\right)\right)+r^6 \left(\frac{r_0}{r}\right)^{\eta +1} \left(2 (\eta -3) \mu +24 \pi  (\eta +1)+\lambda  \phi _0 \left(\frac{r_0}{r}\right)^{\lambda } \left(\mu  (\eta +6 \lambda +37)
       \right.\right.\right.\\ \left.\left.\left.
     \hspace{0.4cm} -24 \pi  (\eta +6 \lambda -1)-6 \lambda  (24 \pi -\mu ) \phi _0 \left(\frac{r_0}{r}\right)^{\lambda }\right)\right)+2 \lambda  r^6 \phi _0 \left(\frac{r_0}{r}\right)^{\lambda } \left(-(\lambda +5) \mu +24 \pi  \lambda +\lambda  (24 \pi -\mu ) \phi _0 \left(\frac{r_0}{r}\right)^{\lambda }\right)+\gamma  
\right.\\ \left.
     \hspace{0.3cm} r_0^4 \left(\frac{r_0}{r}\right)^{4 \eta } \left(-(\eta  (\eta +18)+33) \mu -24 \pi  (\eta +1)^2+2 \lambda  \phi _0 \left(\frac{r_0}{r}\right)^{\lambda } \left(-\mu  (\eta  (\eta +2 \lambda +22)+2 \lambda +23)+24 \pi  (\eta  (\eta +2 \lambda +4)
     \right.\right.\right.\\ \left.\left.\left.
    +2 \lambda +5)+2 (\eta +1) \lambda  (24 \pi -\mu )\phi _0 \left(\frac{r_0}{r}\right)^{\lambda }\right)\right)\right] \,.
\end{multline}
Finding the null energy condition (NEC) in the radial and tangential directions is made possible by the following components
\begin{equation}\label{83}
    \rho+p_r = -\frac{\left(r_0 \left(\frac{r_0}{r}\right)^{\eta } \left(2 \gamma  (\eta +1) r_0 \left(\frac{r_0}{r}\right)^{\eta }-r^3\right)+r^4\right) \left(r_0 \left(\frac{r_0}{r}\right)^{\eta } \left(\eta -2 \lambda  \phi _0 \left(\frac{r_0}{r}\right)^{\lambda }+1\right)+2 \lambda  r \phi _0 \left(\frac{r_0}{r}\right)^{\lambda }\right)}{(\mu +8 \pi ) r^6 \left(r-r_0 \left(\frac{r_0}{r}\right)^{\eta }\right)} \,,
\end{equation}
\begin{multline}\label{84}
    \rho+p_t =  \frac{1}{2 (\mu +8 \pi ) r^6 \left(r-r_0 \left(\frac{r_0}{r}\right)^{\eta }\right)^2} \left[2 r^2 r_0^2 \left(\frac{r_0}{r}\right)^{2 \eta } \left(\lambda  \phi _0 \left(\frac{r_0}{r}\right){}^{\lambda } \left(2 \gamma  (\eta  (\lambda +3)+\lambda +4)+\lambda  \phi _0 \left(\frac{r_0}{r}\right)^{\lambda } \left(2 \gamma  (\eta +1)+
    \right.\right.\right.\right. \\ \left.\left.\left.\left.
   \hspace{1.4cm} 3 r^2\right)+r^2 (\eta +3 \lambda +1)\right)+(\eta -1) r^2\right)-r r_0^3 \left(\frac{r_0}{r}\right)^{3 \eta } \left(2 \gamma  (\eta  (\eta +6)+7)+\lambda  \phi _0 \left(\frac{r_0}{r}\right)^{\lambda } \left(2 \gamma  (\eta  (\eta +4 \lambda +12)+4 \lambda 
    \right.\right.\right. \\ \left.\left.\left.
  \hspace{1.4cm}  +15)+2 \lambda  \phi _0 \left(\frac{r_0}{r}\right)^{\lambda } \left(4 \gamma  (\eta +1)+r^2\right)+r^2 (\eta +2 \lambda +1)\right)+(\eta -1) r^2\right)-r^6 \left(\frac{r_0}{r}\right)^{\eta +1} \left(\eta +\lambda  \phi _0 \left(\frac{r_0}{r}\right)^{\lambda } \left(\eta +6 \lambda 
    \right.\right.\right. \\ \left.\left.\left.
   \hspace{1.2cm} +6 \lambda  \phi _0 \left(\frac{r_0}{r}\right)^{\lambda }+1\right)-1\right)+2 \lambda ^2 r^6 \phi _0 \left(\frac{r_0}{r}\right)^{\lambda } \left(\phi _0 \left(\frac{r_0}{r}\right)^{\lambda }+1\right)+2 \gamma  r_0^4 \left(\frac{r_0}{r}\right)^{4 \eta } \left(\eta  (\eta +4)+\lambda  \phi _0 \left(\frac{r_0}{r}\right)^{\lambda } \left(\eta  (\eta +2 \lambda
    \right.\right.\right. \\ \left.\left.\left.
    +6) +2 \lambda +2 (\eta +1) \lambda  \phi _0 \left(\frac{r_0}{r}\right)^{\lambda }+7\right)+5\right)\right] \,.
\end{multline}
In this particular instance, we observe that the NEC along the radial and tangential directions become undefinable at the wormhole's throat, or $r=r_0$. This demonstrates that wormhole solutions are impossible to achieve using this shape function \eqref{16}. As a result, we draw the conclusion that postulating a non-linear form \eqref{79} is inappropriate for wormhole solutions with the shape function \eqref{16}. Yet there are also other options for shape function that we might explore more in the future.

\section{Junction condition}\label{sec6}
As we know that there are two different metrics across the thin shell to match the condition along the boundary, we need to use the Israel junction condition to get the solutions (In general, we use the junction condition for the fact that along the hypersurfaces the metric has to be continuous as well as differentiable so we check both the Christoffel symbol as well the Riemann curvature tensor to see what the boundary condition leads to, as in thin shell formulation we can get such thing via following the calculation done below.).\\
 We also note that the thin shell or the three-manifold would be denoted by $\Sigma$, outside of the thin shell there are Schwarzschild solutions denoted by $M^+$, and inside there is wormhole $M^-$ the total space-time would be $M^+\cup\Sigma\cup M^-$. We would focus stress $\sigma$ and pressure $p$ and we would construct the effective potential from that as well see the condition for given exotic matter given the null energy condition violation. \\
 This can be done as follows we have interior solutions as,
 \begin{equation}\label{85}
ds^2=e^{2\phi(r)}dt^2-(f(r))^{-1}dr^2-r^2\,d\theta^2-r^2\,\sin^2\theta\,d\Phi^2\,.
\end{equation}
 For the exterior solutions, we would take the Schwarzschild  solution of the form,
 \begin{equation}\label{86}
     ds^2=-F(r)dt^2+(F(r))^{-1}dr^2+r^2+d\Omega^2\,.
 \end{equation}
We also note that in both cases the space-like component is spherically symmetric, on the boundary we would get the FLRW metric,
 \begin{equation}\label{87}
     ds^2=-d\tau^2+a(\tau)^2 d\Omega^2\,.
 \end{equation}
 Now, if we take the formula for the first junction condition we get,
 \begin{equation}\label{88}
     K_{ab}^{\pm}=-n_{\gamma}^{\pm}\left(\frac{\partial^2x^{\gamma}_{\pm}}{\partial \zeta^a \partial \zeta^b}+\Gamma^{\gamma}_{\alpha\beta}\frac{\partial x^{\alpha}_{\pm}}{d\zeta^a}\frac{\partial x^{\beta}_{\pm}}{d\zeta^b}\right)\,,
 \end{equation}
 using the first Israel junction condition we will get the proper boundary condition.\\
 For interior geometry, we get the following components,
 \begin{equation}
 K^{\tau-}_{\tau}=\frac{f'(a)+2\Ddot{a}}{2\sqrt{f(a)+\dot{a}}};\,\,\,\,\,\,\,\,\,\,\,\,\,\,K^{\theta-}_{\theta}=\frac{\sqrt{f(a)+\dot{a}^2}}{a};\,\,\,\,\,\,\,\,\,\,\,\,\,\, K^{\phi-}_{\phi}=\sin^2{\theta} K^{\theta-}_{\theta}\,.
 \end{equation}
 For exterior geometry, we get the following components,
 \begin{equation}\label{89}
K^{\tau+}_{\tau}=\frac{F'(a)+2\Ddot{a}}{2\sqrt{F(a)+\dot{a}}};\,\,\,\,\,\,\,\,\,\,\,\,\,\,K^{\theta+}_{\theta}=\frac{\sqrt{F(a)+\dot{a}^2}}{a};\,\,\,\,\,\,\,\,\,\,\,\,\,\,K^{\phi+}_{\phi}=\sin^2{\theta} K^{\theta+}_{\theta}\,,
 \end{equation}
 where dot denotes the derivative with respect to the proper time ($\tau$) and prime denotes the derivative with respect to ordinary time.\\
 Now, to calculate the surface stress and pressure for the thin shell to sustain itself we use the Lankoz equation,
 \begin{equation}\label{90}
     S_{ab}=\frac{1}{8\pi}\left[g_{ab}K-K_{ab}\right]\,,
 \end{equation}
 where $a,b=0,2,3$ as we have discussed at the shell $r$ is constant,
 \begin{equation}\label{91}
     \sigma(a)=-\frac{1}{4\pi a}\left[\sqrt{F(a)+\dot{a}^2}-\sqrt{f(a)+\dot{a}^2}\right]
 \end{equation}
 and
 \begin{equation}\label{92}
     p(a)=\frac{1}{16\pi a}\left[\frac{2F(a)+aF'(a)+2a\Ddot{a}+2\dot{a}^2}{\sqrt{F(a)+\dot{a}^2}}-\frac{2f(a)+af'(a)+2a\Ddot{a}+2\dot{a}^2}{\sqrt{f(a)+\dot{a}^2}}\right]\,.
 \end{equation}
 We note that at throat $a=a_0$ we get $\dot{a_0}=0$,
 so at the throat, we get 
 \begin{equation}\label{93}
     \sigma(a_0)=-\frac{1}{4\pi a_0}\left[\sqrt{F(a_0)}-\sqrt{f(a_0)}\right]
 \end{equation}
 and also,
  \begin{equation}\label{94}
     p(a_0)=\frac{1}{16\pi a_0}\left[\frac{2F(a_0)+a_0F'(a_0)}{\sqrt{F(a_0)}}-\frac{2f(a_0)+a_0f'(a_0)}{\sqrt{f(a_0)}}\right]\,.
 \end{equation}
We note that at the throat the NEC needs to be violated that is $\sigma(a_0)+p(a_0)<0$ at the throat. The violation of the null energy condition on the shell would imply the presence of exotic matter.\\
We note that by following the prescription given by \cite{M. Sharif, S. D. Forghani}. We can also go further and can calculate the potential $V(r)$ by noting that the energy-momentum has a conservation relation:
\begin{equation}\label{95}
    \frac{d}{d\tau}(\sigma \phi)=p\frac{d\phi}{d\tau}=0\,,
\end{equation}
where $\sigma=4\pi a^2$. From the conservation equation above one can find 
\begin{equation}\label{96}
    \sigma '=-\frac{2}{a}(\sigma+p)\,.
\end{equation}
Following the prescription given in \cite{E. Poisson}, we note that the last equation has the form of $\dot{a}^2+V(a)$, so from the above equation we can get
\begin{equation}\label{97}
    V(a)=\frac{f(a)}{2}+\frac{F(a)}{2}-\frac{(f(a)-F(a))^2}{64a^2\pi^2\sigma^2}-4a^2\pi^2\sigma^2\,.
\end{equation}
We also note that in our case $f(r)$ and $F(r)$ are given by the following, 
 $$f(r)=1-\frac{b(r)}{r}$$ and $$F(r)=1-\frac{2GM}{r}$$\,.

We note that outside the thin shell, we can take the Schwarzschild solution because there is no matter so the Schwarzschild solution is applicable in a vacuum with spherical symmetry.
\begin{figure}[H]
\centering
\includegraphics[width=15.5cm,height=4cm]{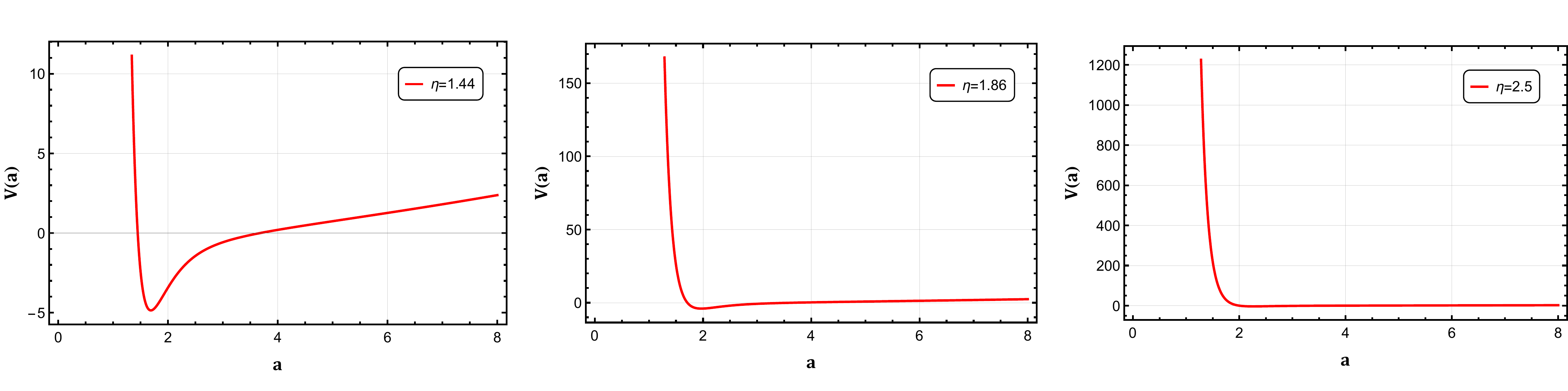}
\caption{Profile shows the behaviour of $V(a)$ for $M =0.2$ and $a_0 =3.68$\,.}
\label{fig3}
\end{figure}
The above figure shows that a  thin shell in the junction feels a similar potential (in natural units) to that of a massive particle in the Schwarzschild metric.

\section{Final Remarks}
\label{sec7}
The present study involves a comprehensive analytical investigation of the parameter space for a particular family of wormhole solutions within the framework of the linear $f(Q, T) = \alpha Q + \beta T$ theory of gravity. The main objective is to determine the necessary constraints on the free parameters of the model that ensure the traversability and non-exoticity of the wormhole solutions, which implies that all energy conditions are satisfied across the entire spacetime. Additionally, we demonstrate that even in cases where some parameter bounds are exceeded, resulting in the wormhole becoming exotic beyond the throat at a finite radius $r_c$, such exoticity can be effectively eliminated through a spacetime matching to an exterior vacuum solution.

We have considered a family of wormholes with redshift and shape functions described by the expressions in Eqs. \eqref{15} and \eqref{16}, respectively, within a linear version of $f(Q, T) = \alpha Q + \beta T$ theory. Our main finding is that ensuring the wormhole solution satisfies the NEC for the entire spacetime automatically guarantees that the WEC and SEC are also satisfied. This is due to the fact that the parameter bounds arising from the energy conditions $\rho > 0$ and $\rho + p_r + 2p_t > 0$ are weaker than those arising from $\rho + p_i > 0$. However, it should be noted that the implications are one-directional, and thus a solution satisfying $\rho > 0$ or $\rho + p_r + 2p_t > 0$ may not necessarily satisfy the NEC for the whole spacetime. Regarding the DEC, we have shown that the bounds arising from $\rho-p_i > 0$ are stronger than those from $\rho + p_i > 0$, which means that a solution satisfying the NEC may not necessarily satisfy the DEC. Nonetheless, we have demonstrated that strong solutions satisfying all four energy conditions (NEC, WEC, SEC, and DEC) can be achieved for a wide range of parameter combinations. Our choice of the linear form $f(Q, T)$ enables us to perform an analytical study of the parameter space and to prove that even the simplest extension of GR within the framework of $f(R, T)$ can effectively address the issue of exotic matter in wormhole spacetimes.

In the article, we have also shown the stability of a thin-shell around the wormhole. We have calculated the stress ($\sigma$) and pressure ($P$) of such a thin shell. We have also shown that what are the conditions under which the thin-shell will have exotic matter (by checking the null energy condition.). We have also shown that using the energy-momentum conservation equation one can find the potential ($V(a)$) across the thin shell. We have also drawn the shape of $V$ for various values of $\eta$ and have shown that it has a similar behavior as a particle feels outside the Schwarzchild radius. One can use this potential to calculate the gravitational lensing, accretion disk (via ISCO), etc for such a thin shell. From the observation of gravitational lensing, phenomena around a thin shell made out of dust. one can reconstruct the potential and can get shape function, etc by backward bootstrap.

\section*{Data Availability Statement}
There are no new data associated with this article.

 \acknowledgements  MT acknowledges University Grants Commission (UGC), New Delhi, India, for awarding National Fellowship for Scheduled Caste Students (UGC-Ref. No.: 201610123801). SG acknowledges Council of Scientific and Industrial Research (CSIR), Government of India, New Delhi, for junior research fellowship (File no.09/1026(13105)/2022-EMR-I). PKS acknowledges National Board for Higher Mathematics (NBHM) under Department of Atomic Energy (DAE), Govt. of India for financial support to carry out the Research project No.: 02011/3/2022 NBHM(R.P.)/R\&D II/2152 Dt.14.02.2022. We are very much grateful to the honorable referee and to the editor for the illuminating suggestions that have significantly improved our work in terms
of research quality, and presentation.

\end{document}